\documentclass[useAMS]{mn2e}

\usepackage{times}
\usepackage{graphicx}
\usepackage{amsmath}
\usepackage{amssymb}
\newcommand{\beq}{\begin{equation}}
\newcommand{\bea}{\begin{eqnarray}}
\newcommand{\eeq}{\end{equation}}
\newcommand{\eea}{\end{eqnarray}}
\newcommand{\im}{{\cal I}}
\newcommand{\bfs}[1]{\mbox{\boldmath{$#1$}}}

\title[Instabilities at ultra-relativistic shock waves]{On
  electromagnetic instabilities at ultra-relativistic shock waves}

\author[M. Lemoine and G. Pelletier]
{Martin Lemoine$^{1}$\thanks{e-mail:{\tt lemoine@iap.fr}} and
Guy Pelletier$^{2}$\thanks{e-mail:{\tt
     guy.pelletier@obs.ujf-grenoble.fr}}\\
$^{1}$ Institut d'Astrophysique de Paris, \\ 
	CNRS, UPMC, 
	98 bis boulevard Arago, F-75014 Paris, France\\
$^{2}$ Laboratoire d'Astrophysique de Grenoble, \\
	CNRS, Universit\'e Joseph Fourier II,
	BP 53, F-38041 Grenoble, France; \\
}

\begin{document}

\date{}

\pubyear{2008}

\maketitle

\label{firstpage}

\begin{abstract}
  Recent work on Fermi acceleration at ultra-relativistic shock waves
  has demonstrated the need for strong amplification of the background
  magnetic field on very short scales. Amplification of the magnetic
  field by several orders of magnitude has also been suggested by
  observations of gamma-ray bursts afterglows, both in downstream and
  upstream plasmas. This paper addresses this issue of magnetic field
  generation in a relativistic shock precursor through
  micro-instabilities. In a generic superluminal configuration, the
  level of magnetization of the upstream plasma turns out to be a
  crucial parameter, notably because the length scale of the shock
  precursor is limited by the Larmor rotation of the accelerated
  particles in the background magnetic field and by the speed of the
  shock wave. We discuss in detail and calculate the growth rates of
  the following beam plasma instabilities seeded by the accelerated
  and reflected particle populations: for an unmagnetized shock, the
  Weibel and filamentation instabilities, as well as the \v{C}erenkov
  resonant instabilities with electrostatic modes; for a magnetized
  shock, the Weibel instability and the resonant \v{C}erenkov
  instabilities with the longitudinal electrostatic modes, as well as
  the Alfv\'en, Whisler and extraordinary modes. All these
  instabilities are generated upstream, then they are transmitted
  downstream. The modes excited by \v{C}erenkov resonant instabilities
  take on particular importance with respect to the magnetisation of
  the downstream medium since, being plasma eigenmodes, they have a
  longer lifetime than the Weibel modes. We discuss the main
  limitation of the wave growth associated with the length of
  precursor and the magnetisation of the upstream medium for both
  oblique and parallel relativistic shock waves. We also characterize
  the proper conditions to obtain Fermi acceleration at
  ultra-relativistic shock waves: for superluminal shock waves, the
  Fermi process works for values of the magnetization parameter below
  some critical value, and there is an intrinsic limitation of the
  achievable cosmic ray energy depending on the ratio of the
  magnetization to its critical value. We recover results of most
  recent particle-in-cell simulations and conclude with some
  applications to astrophysical cases of interest. In particular,
  Fermi acceleration in pulsar winds is found to be unlikely whereas
  its development appears to hinge on the level of upstream
  magnetization in the case of ultra-relativistic gamma-ray burst
  external shock waves.
\end{abstract}


\section{Introduction}
Substantial progress has been accomplished in this last decade on our
theoretical understanding of the acceleration of particles at
relativistic shocks, revealing in more than one place crucial
differences with Fermi acceleration at non-relativistic shock
waves. For instance, Gallant \& Achterberg (1999), Achterberg et
al. (2001) have emphasized the strong anisotropy of the cosmic ray
population propagating upstream, which is directly related to the fact
that the relativistic shock wave is always trailing right behind the
accelerated particles. These particles are confined into a beam of
opening angle $\theta\,\lesssim\,1/\Gamma_{\rm sh}$ (with $\Gamma_{\rm
  sh}$ the Lorentz factor of the shock wave in the upstream frame) and
are overtaken by the shock wave on a timescale $r_{\rm L}/\Gamma_{\rm
  sh}$, with $r_{\rm L}$ the typical Larmor radius of these particles
in the background magnetic field. One consequence of the above is to
restrict the energy gain per
up$\,\rightarrow\,$down$\,\rightarrow\,$up cycle, $\Delta E/E$, to a
factor of order unity.  Early Monte Carlo numerical experiments
nonetheless observed efficient Fermi acceleration, with a generic
spectral index $s=2.2-2.3$ in the ultra-relativistic limit (Bednarz \&
Ostrowski 1998, Achterberg et al. 2001, Lemoine \& Pelletier 2003,
Ellison \& Double 2004), in agreement with semi-analytical studies
(Kirk et al. 2000) and analytical calculations (Keshet \& Waxman
2005). This value of the spectral index is however restricted to the
assumption of isotropic turbulence both upstream and downstream of the
shock (Niemiec \& Ostrowski 2004; Lemoine \& Revenu 2006), whereas the
shock crossing conditions imply a mostly perpendicular magnetic field
downstream, which severely limits the possibility of downstream
scattering. Furthermore, it was later stressed by Niemiec \& Ostrowski
(2006) and Lemoine, Pelletier \& Revenu (2006) that these early
studies implicitly ignored the correlation between the upstream and
downstream particle trajectories during a cycle. In particular, the
former numerical study demonstrated that Fermi acceleration became
inefficient if the proper shock crossing conditions were applied to
the background magnetic field. This result was demonstrated
analytically in the latter study, concluding that Fermi acceleration
could only proceed if strong turbulence ($\delta B \gg B$) existed on
a scale much smaller than the typical larmor radius. The addition of
turbulence on large scales $\,\gg\,r_{\rm L}$ does not help in this
respect, as the particle then experiences a roughly coherent field on
the short length scales that it probes during its cycle. Further
studies by Niemiec, Ostrowski \& Pohl (2006) have confirmed that Fermi
acceleration proceeds if short scale turbulence is excited to high
levels, either downstream or upstream. The detailed conditions under
which Fermi acceleration can proceed have been discussed analytically
in Pelletier, Lemoine \& Marcowith (2009); they are found to agree
with the numerical results of Niemiec, Ostrowski \& Pohl (2006).

Amplification of magnetic fields on short spatial scales thus appears
to be an essential ingredient in Fermi processes at ultra-relativistic
shock waves. Quite interestingly, strong amplification has been
inferred from the synchrotron interpretation of gamma-ray burst
afterglows, downstream at the level of $\delta B/B \,\gtrsim\,
10^4-10^5$ (Waxman 1997; see Piran 2005 for a review), and upstream
with $\delta B/B \,\gtrsim\, 10^2-10^3$ (Li \& Waxman 2006), assuming
an upstream magnetic field typical of the interstellar
medium. Understanding the mechanism by which the magnetic field gets
amplified is crucial to our understanding to relativistic Fermi
acceleration, since the nature of this short scale turbulence will
eventually determine the nature of scattering, hence the spectral
index and the acceleration timescale.

Concerning the amplification of the downstream magnetic field, the
Weibel two stream instability operating in the shock transition layer
has been considered as a prime suspect (Gruzinov \& Waxman 1999,
Medvedev \& Loeb 1999, Wiersma \& Achterberg 2004, Achterberg \&
Wiersma 2007, Achterberg, Wiersma \& Norman 2007; Lyubarsky \& Eichler
2006). Several questions nevertheless remain open. For instance,
Hededal \& Nishikawa (2005) and Spitkovsky (2005) have observed, by
the means of numerical simulations that this instability gets quenched
when the magnetization of the upstream field becomes sufficiently
large. On analytical grounds, Wiersma \& Achterberg (2004), Achterberg
\& Wiersma (2007), and Lyubarsky \& Eichler (2006) have argued that it
saturates at a level too low to explain the gamma-ray burst
afterglow. The long term evolution of the generated turbulence also
remains an open question, although Medvedev et al. (2005) claim to see
the merging of current filaments into larger filaments through
dedicated numerical experiments.

Regarding upstream instabilities, the relativistic generalization of
the non-resonant Bell instability has been investigated by
Milosavljevi\'c \& Nakar (2006) and Reville, Kirk \& Duffy (2006) in the
case of parallel shock waves. However, ultra-relativistic shock waves
are generically superluminal, with an essentially transverse magnetic
field in the shock front. For this latter case, Pelletier, Lemoine \&
Marcowith (2009) have shown that the equivalent of the Bell
non-resonant instability excites magnetosonic compressive modes and
saturates at a moderate level $\delta B/ B\,\sim\,1$ in the frame of
the linear theory.

In recent years, particle-in-cell (PIC) simulations have become a key
tool in the investigation of these various issues. Such simulations go
(by construction) beyond the test particle approximation and may
therefore probe the wave -- particle relationship, which is central to
all of the above issues. Of course, such benefice comes at the price
of numerical limitations of the simulations, both in terms of
dimensionality and of dynamic range, which in turns impact on the mass
ratios accessible to the computation. Nonetheless, early PIC
simulations have been able to simulate the interpenetration of
relativistic flows and to study the development of two stream
instabilities at early times, see e.g. Silva et al. (2003),
Frederiksen et al. (2004), Hededal et al. (2004), Dieckmann (2005),
Dieckmann, Drury \& Shukla (2006), Dieckmann, Shukla \& Drury (2006),
Nishikawa et al. (2006), Nishikawa et al. (2007) and Frederiksen \&
Dieckmann (2008) for unmagnetized colliding plasma shells, and
Nishikawa et al. (2003), Dieckmann, Eliasson \& Shukla (2004a, b),
Nishikawa et al. (2005) and Hededal \& Nishikawa (2005) for studies of
the magnetized case. The formation of the shock itself has been
observed for both electron-positron and electron-proton plasmas thanks
to recent simulations that were able to carry the integration on to
longer timescales, see e.g. Spitkovsky (2005), Kato (2007), Chang et
al. (2008), Dieckmann, Shukla \& Drury (2008), Spitkovsky (2008a, b),
Keshet et al. (2009).  All of the above studies use different
techniques for the numerical integration, and varying parameters
(dimensions, composition, mass ratios, density ratios of the colliding
plasmas and relative Lorentz factors) in order to examine different
aspects of the instabilities to various degrees of accuracy and over
different timescales.

Several of these studies have reported hints for particle acceleration
through non Fermi processes (Dieckmann, Eliasson \& Shukla 2004b;
Frederiksen et al. 2004; Hededal et al. 2004; Hededal \& Nishikawa
2005; Nishikawa et al. 2005; Dieckmann, Shukla \& Drury 2006,
2008). Concrete evidence for Fermi acceleration, i.e. particles
bouncing back and forth across the shock wave has come with the recent
simulations of Spitkovsky (2008b), and was studied in more details for
both magnetized and unmagnetized shock waves in Sironi \& Spitkovsky
(2009). In particular, this latter study has demonstrated the
inefficiency of Fermi acceleration at high upstream magnetization in
the superluminal case, along with the absence of amplification of the
magnetic field (thus in full agreement with the calculations of
Lemoine, Pelletier \& Revenu 2006). This result is particularly
interesting, because it suggests that the magnetization of the
upstream plasma, in limiting the length of the precursor, may hamper
the growth of small scale magnetic fields, and therefore inhibit Fermi
cycles. Finally, the long term simulations of Keshet et al. (2009)
have also observed a steady development of turbulence upstream of the
shock wave, suggesting that as time proceeds, particles are
accelerated to higher and higher energies and may thus stream further
ahead of the shock wave. We will discuss this issue as well at the end
of the present work.

The main objective of this paper is to undertake a systematic study of
micro-instabilities in the upstream medium of a relativistic shock
wave. We should emphasize that we assume the shock structure to exist
and we concentrate our study on the shock transition region where the
incoming upstream plasma collides with the shock reflected and shock
accelerated ions that are moving towards upstream infinity. Therefore,
care should be taken when confronting the present results to the above
numerical simulations which reproduce the collision of two neutral
plasma flows in order to study the development of instabilities that
eventually lead to the formation of the shock (through the
thermalization of the electron and ion populations).  The physical
set-up that we have in mind matches best that obtained in the
simulations of shock formation and particle acceleration described in
Spitkovsky (2008b), Keshet et al. (2009) and Sironi \& Spitkovsky
(2009), or that simulated in Dieckmann, Eliasson and Shukla (2004a, b)
and Frederiksen \& Dieckmann (2008), or that studied in Medvedev \&
Zakutnyaya (2008). Our approach also rests on the following
observation, namely that in the ultra-relativistic limit, the
accelerated (or the reflected) particle population essentially behaves
as an {\em unmagnetized cold beam of Lorentz factor $\sim\,\Gamma_{\rm
    sh}^2$}.  

In the present paper, we assume the beam to be carrying a weak current
and in so doing, we neglect electromagnetic current instabilities. We
will nevertheless include in our summary of instabilities the
relativistic generalization of the Bell current instability (Bell
2004), since it has been studied in detail in several recent studies
(Milosavljevi\'c \& Nakar 2006; Reville, Kirk \& Duffy 2006). The
instability triggered by the cosmic-ray current in the case of oblique
shock waves has also been discussed in the relativistic regime in
Pelletier, Lemoine \& Marcowith (2009). Note also that in the case of
pair plasmas, electromagnetic current instabilities do not take place
as the beam remains neutral.

The layout of the present paper is as follows. We examine the
instabilities triggered by this beam, considering in turn the cases of
an unmagnetized upstream plasma (Section~3) and that of a magnetized
plasma (Section~4). In Section~5, we discuss the intermediate limit
and construct a phase diagram indicating which instability prevails as
a function of shock Lorentz factor and magnetization level. We then
discuss the possibility of Fermi acceleration in the generated
turbulence and apply these results to the case of gamma-ray bursts
shock waves and pulsar winds. We will recover the trend announced
above, namely that a magnetized upstream medium inhibits the growth of
the magnetic field hence particle acceleration. In Section~2, we first
discuss the general structure of a collisionless shock, in the case of
a electron--proton plasma with a quasi perpendicular mean field,
borrowing from analyses in the non-relativistic limit.

\section{General considerations}

\subsection{On the configuration of a relativistic collisionless shock wave}

A collisionless shock is built with the reflection of a fraction of
incoming particles at some barrier, generally of electrostatic or
magnetic nature. Let us sketch the general picture, borrowing from
model of non-relativistic collisionless electro-ion shocks (see
e.g. Treumann \& Jaroschek 2008a, b for recent reviews).  In an
electron--proton- plasma carrying an oblique magnetic field, one
expects a barrier of both electrostatic and magnetic nature to
rise. Because the magnetic field is frozen in most part of the plasma,
its transverse component is amplified by the velocity decrease. This
in itself forms a magnetic barrier which can reflect back a fraction
of the incoming protons.  Similarly, the increase of electron density
together with the approach of the electron population towards
statistical equilibrium is concomittant with the rise of an
electrostatic potential such that $e\Phi \simeq T_e \log (n/n_{\rm
  u\vert\rm sh})$ ($n$ is the local density in the front frame, and
$n_{\rm u\vert\rm sh}$ the upstream incoming density viewed in the
front frame). The electron temperature is expected to grow to a value
comparable to, but likely different from that of protons, which
reaches $T_p \,\sim\, \left(\Gamma_{\rm sh}-1\right)m_pc^2$. The
electrostatic barrier thus allows the reflection of a significant part
of the incoming protons since $e \Phi \,\sim\, \left(\Gamma_{\rm
  sh}-1\right)m_pc^2$. Although it reflects a fraction of protons, it
favors the transmission of electrons that would otherwise be reflected
by the magnetic barrier. The reflection of a fraction of the protons
ensures the matter flux preservation against the mass density increase
downstream. However because the magnetic field is almost transverse,
an intense electric field $E \,=\,\beta_{\rm sh} B$ energizes these
reflected protons such that they eventually cross the
barrier. Interactions between the different streams of protons are
then expected to generate a turbulent heating of the proton
population, which takes place mostly in the so-called ``foot''
region. This foot region extends from the barrier upstream over a
length scale (in the shock front frame, as indicated by the
$_{\vert\rm sh}$ subscript) $\ell_{\rm F\vert\rm sh} = r_{\rm L\vert
  sh}$, where $r_{\rm L\vert\rm sh}$ denotes the Larmor radius of the
reflected protons.

Entropy production in the shock transition region comes from two
independent anomalous (caused by collisionless effects) heating
processes for electrons and ions. The three ion beams in the foot
(incoming, reflected in the foot and accelerated) interact through the
``modified two stream instability'', which seemingly constitutes the
main thermalisation process of the ion population. A careful
description of these anomalous heating processes certainly requires an
appropriate kinetic description. For the time being, we note that the
growth of the ion temperature develops on a length scale $\ell_{\rm
  F}$. The temperature of the electrons rather grows on a very short
scale scale $\ell_{\rm R}\,\ll\,\ell_{\rm F}$ which defines the
``ramp'' of the shock.  In non-relativistic shocks, electrons reach a
temperature larger than ions; however we do not know yet whether this
is the case in relativistic shocks. These electrons also experience
heating in the convection electric field.  Moreover, due to the strong
gradient of magnetic field, an intense transverse electric current is
concentrated, inducing anomalous heat transfer through the
ramp. Probably an anomalous diffusion of electron temperature occurs
that smoothes out the temperature profile; however it has not been
identified in relativistic shocks. Electron heating is described by
Ohm's law in the direction of the convection electric field (in the
$\mathbf{x\times B}$ direction, taken to be $\mathbf{z}$):
\begin{equation}
\label{eq:jh}
\beta_x B + E = \frac{\eta c}{4\pi} \frac{{\rm d}B}{{\rm d}x}\ ,
\end{equation}
with $\beta_x<0$ in the shock front frame, $E\,=\,\beta_{\rm sh}B_{\rm
  u}$, $B_{\rm u}$ denoting the background magnetic field at
infinity. The magnetic field profile can be obtained by prescribing a
velocity profile going from $-\beta_{\rm sh}c\sim -c$ to $\simeq-c/3$
over a distance much larger than $\ell_{\rm R}$. The profile displays
a ramp at scale $\ell_{\rm R}$ followed by an overshoot before
reaching the asymptotic value $3B_{\rm u}$. The above result indicates that
the relevant scale for $\ell_{\rm R}$ is the relativistic resistive
length:
\begin{equation}
\ell_{\rm R} \sim \frac{\eta c}{4\pi} = \delta_{\rm e} 
\frac{\nu_{\rm eff}}{\omega_{\rm pe}}\ .
\end{equation}
This is a very short scale not larger than the electron inertial
length $\delta_{\rm e}\,\equiv\,c/\omega_{\rm pe}$ even when the anomalous
resistivity is so strong that the effective collision frequency
$\nu_{\rm eff}$ is of order $\omega_{\rm pe}$. This scale thus represents the
growth scale of three major quantities, namely, the potential, the
magnetic field and the electron temperature. It is of interest to
point out that this scale always remains much smaller than the foot
scale. Indeed, even if $\delta_{\rm e}$ is estimated with ultra-relativistic
electrons of relativistic mass $\Gamma_{\rm sh} m_p$,
i.e. $\delta_{\rm e}\,=\, \left[\Gamma_{\rm sh} m_p c^2/ (4\pi n_{e\vert\rm
    sh} e^2)\right]^{1/2}$, it remains smaller than the foot length,
since
\begin{equation}
\label{eq:2}
\frac{\delta_{\rm e}}{\ell_{\rm F\vert\rm sh}}\,=\, \left(\frac{B_{\vert\rm
      sh}^2}
  {4\pi n_{e\vert\rm u} \Gamma_{\rm sh}^2 m_p c^2}\right)^{1/2} \,\ll\,1 \ ,
\end{equation}
using the value of $\ell_{\rm F\vert\rm sh}$ for particles with typical
energy $\Gamma_{\rm sh}m_pc^2$ in the shock front. The last inequality
in the above equation is a natural requirement for a strong shock.
The downstream flow results from the mixing of the flow of first
crossing ions (adiabatically slowed down) with the flow of transmitted
ions after reflection. All the ingredients of a shock are then
realized.

In the case of an electron-positron plasma, when a magnetic field is
considered, no electrostatic barrier rises, only the magnetic barrier
appears. However, if the mean magnetic field is negligible, a barrier
can rise only through the excitation of waves, as demonstrated by the
PIC simulations discussed above.

The structure is thus described by two scales $\ell_{\rm R}$ and
$\ell_{\rm F}$ and three small parameters: $\xi_{\rm cr}$, the
fraction of thermal energy density behind the shock converted into
cosmic ray energy, $\sigma_B$ the ratio of magnetic energy density
over the incoming energy density and $1/ \Gamma_{\rm sh}$.

\subsection{Particle motion}
As mentioned above, there are three particle populations in the foot:
the cold incoming particles, the reflected protons, and the
accelerated particle population which has undergone at least one
up$\,\rightarrow\,$down$\,\rightarrow\,$up cycle. This latter
population arrives upstream with a typical Lorentz factor
$\Gamma_{\rm b}\,\sim\,\Gamma_{\rm sh}^2$, with a typical relative spread
of order unity. The second population of reflected protons also
carries an energy $\simeq\Gamma_{\rm sh}^2m_pc^2$, since these
particles have performed a Fermi-like cycle, albeit in the front
rather than downstream.  Therefore one can treat these two populations
as a single beam. From the point of view of the instabilities, one can
approximate this beam as cold, with momentum distribution $\propto
\delta\left(p_x-\Gamma_{\rm
  sh}^2m_pc\right)\delta\left(p_\perp\right)$. Indeed, the
instabilities are governed by the beam velocity, the dispersion of
which remains very small, being of order
$\Delta\beta_{\rm b}\,\sim\,-(2/\Gamma_{\rm b}^2)\Delta\Gamma_{\rm b}/\Gamma_{\rm b}$.
In order to verify this, one writes the susceptibility of the beam,
assuming as above that it is unmagnetized on the scale of the
instabilities (Melrose 1986):
\begin{eqnarray}
\chi^{\rm b}_{ij}&\,=\,&-\frac{4\pi e^2}{m_p \omega^2}\int \frac{{\rm
    d}^3p}{\gamma}\,
f_{\rm b}(\mathbf{p})\nonumber\\ &&\,\,\,\times\left[\delta_{ij}+ \frac{k_i
    c\beta_{j}+ k_jc \beta_{i}}{\omega- \mathbf{k}\cdot
    \bfs{\beta}c} + \frac{(k^2c^2-\omega^2)\beta_{i}\beta_{j}}{\left(\omega -
    \mathbf{k}\cdot\bfs{\beta}c\right)^2}\right] \ ,\label{eq:beams}
\end{eqnarray}
with $p=\beta\gamma m_pc$ and $f_{\rm b}(\mathbf{p})$ the distribution
function of the beam. Since the velocity distribution of the beam is
essentially delta like, one may then indeed approximate the above beam
susceptibility with that of a cold beam; the difference amounts to a
redefinition of the beam plasma frequency by a factor of order unity.

Another crucial length scale in our study is the length scale of the
precursor. As discussed above, this length scale $\ell_{\rm F\vert
  sh}=r_{\rm L\vert\rm sh}$ in the front shock in the case of a
magnetized shock wave. In the upstream frame, this can be rewritten
as:
\begin{equation}
\label{eq:lfb}
\ell_{\rm F \vert\rm u} \,\simeq\, \frac{r_{\rm L\vert\rm u}}{\Gamma_{\rm sh}^3}
\,=\, 
\frac{c}{\omega_{\rm ci} \Gamma_{\rm sh} \sin \theta_B} \,\quad
(B_{\rm u}\neq 0)\ .
\end{equation}
We assume that the field is almost perpendicular in the front frame,
but in the upstream comoving frame we consider its obliquity (angle
$\theta_B$ with respect to the shock normal), assuming that $\sin
\theta_B > 1/\Gamma_{\rm sh}$. The particular case of a parallel shock
wave for which $\Gamma_{\rm sh}\sin \theta_B < 1$ is discussed in
Section~\ref{sec:parshock}; there it will be shown that a fraction
$\left(1-\Gamma_{\rm sh}\theta_B\right)^2$ of the particles that
return upstream may actually propagate to upstream infinity in the
limit of a fully coherent upstream magnetic field, while
Eq.~(\ref{eq:lfb}) remains correct for the rest of the accelerated
particle population. The size of the precursor for the particles that
escape away is eventually given by the level of turbulence ahead of
the shock wave.

In the case of an unmagnetized shock wave, the size of the precursor
is determined by the length traveled by the reflected protons in the
self-generated short scale turbulence. Neglecting for simplicity the
influence of the short scale upstream electric fields (we will see in
Section~\ref{sec:Fermi} that this does not affect the following
result), this length scale can be written (Milosavljevi\'c \& Nakar
2006; Pelletier, Lemoine \& Marcowith 2009):
\begin{equation}
\label{eq:lfu}
\ell_{\rm F\vert\rm u}\,\simeq\, \frac{r_{\rm L\vert\rm u}^2}{\Gamma_{\rm
    sh}^4\ell_{\rm c}}
\,\simeq\,\frac{c^2}{\omega_{\rm ci}^2\ell_{\rm c}}\ ,
\end{equation}
where $\ell_{\rm c}$ represents the typical scale of short scale
magnetic fluctuations. Whether one or the other formula applies
depends on several possible situations and outcomes: if the shock is
magnetized and one considers the first generation of cosmic rays, one
should use Eq.~(\ref{eq:lfb}); if the shock is magnetized and one
assumes that a stationary state has developed with strong
self-generated turbulence, one should use Eq.~(\ref{eq:lfu});
obviously, if the development of the turbulence cannot take place, one
should rather use Eq.~(\ref{eq:lfb}); finally, for an unmagnetized
shock, Eq.~(\ref{eq:lfu}) applies. In the following, we discuss the
turbulence growth rate for these different cases.

There seems to be a consensus according to which magnetic fluctuations
have to be tremendously amplified through the generation of cosmic
rays upstream in order for Fermi acceleration to proceed. A fraction
$\xi_{\rm cr}$ of the incoming energy is converted into cosmic rays
and a fraction of this cosmic rays energy is converted into
electromagnetic fluctuations, which add up to a fraction $\xi_{\rm
  em}$ of the incoming energy. This process is expected to develop
such that the generation of cosmic rays allows the generation of
electromagnetic waves that in turn, through more intense scattering,
allows further cosmic ray acceleration and so on until some saturation
occurs. We write the quantities $\xi_{\rm cr}$ and $\xi_{\rm em}$ as:
\begin{equation}
\label{eq:defxi}
\xi_{\rm cr}\,\equiv\, \frac{P_{\rm cr}}{\Gamma_{\rm sh}^2n_{\rm u}m_pc^2}\ ,\quad
\xi_{\rm em}\,\equiv\, \frac{U_{\rm em}}{\Gamma_{\rm sh}^2 n_{\rm u}m_pc^2}\ ,
\end{equation}
with $\xi_{\rm em}\,<\,\xi_{\rm cr}$.  We approximate the beam
pressure with that of the cosmic rays, i.e. $P_{\rm cr}\,\approx\,
\Gamma_{\rm sh}n_{{\rm b}\vert\rm sh}m_p c^2$ for the first generation of
accelerated particles, as expressed in the shock front frame.  The
electromagnetic energy density is written $U_{\rm em}$ in the same
frame, as usual.

Unless otherwise noted, our discussion takes place in the upstream
rest frame in what follows.

\section{Upstream instabilities in the absence of a mean magnetic field}

When the ambient magnetic field can be neglected or is absent, the
reflected particles and the fraction of particles that participate to
the first Fermi cycle constitute a relativistic cold beam that
pervades the ambient plasma and trigger three major
micro-instabilities. One is the two stream electrostatic instability,
which amplifies the electrostatic Langmuir field through a
\v{C}erenkov resonant interaction $\omega - \mathbf{k \cdot v}_{\rm b} =
0$, with
$\mathbf{k}\parallel\mathbf{E}\parallel\mathbf{v}_{\rm b}$. Another is the
Weibel instability, with
$\mathbf{k}\parallel\mathbf{v}_{\rm b}\perp\mathbf{E}$ and its analog
filamentation instability, with
$\mathbf{k}\perp\mathbf{v}_{\rm b}\parallel\mathbf{E}$ (Bret, Firpo \&
Deutsch 2004, 2005a, 2005b; see also Bret 2009 for a recent
compilation). These two instabilities are non-resonant and mostly
electromagnetic with a low phase velocity so that the magnetic
component of the wave is dominant. It is thus particularly relevant
for developing particle scattering. Finally, these authors have also
discovered an oblique resonance which grows faster than the above
two. It is mostly longitudinal (see further below) but $\mathbf{k}$ is
neither perpendicular nor parallel to the beam. These growth rates are
easily recovered as follows.

For a cold beam, Eq.~(\ref{eq:beams}) gives the following
susceptibility:
\begin{equation}
\label{eq:chib}
\chi^{\rm b}_{ij} = -\frac{\omega_{{\rm
      p}{\rm b}}^2}{\omega^2}\Biggl[\delta_{ij} + \frac{k_i c\beta_{{\rm b}j}+
    k_jc \beta_{{\rm b}i}}{\omega- \mathbf{k}\cdot \bfs{\beta}_{\rm b}c} +
  \frac{(k^2c^2-\omega^2)\beta_{{\rm b}i}\beta_{{\rm b}j}}{\left(\omega -
    \mathbf{k}\cdot\bfs{\beta}_{\rm b}c\right)^2}\Biggr] \ .
\end{equation}
The beam propagates with velocity
\mbox{\boldmath{$\beta$}}$_{\rm b}c=\left(1-1/\Gamma_{\rm b}^2\right)^{1/2}\mathbf{x}$;
the relativistic beam plasma frequency (in the upstream frame) is
given by:
\begin{equation}
  \omega_{{\rm p}{\rm b}}\,\equiv\, \left(\frac{4\pi n_{{\rm b}\vert\rm u}e^2}
        {\Gamma_{\rm b}m_p}\right)^{1/2}\ ,
\end{equation}
recalling $\Gamma_{\rm b}\,\simeq\,\Gamma_{\rm sh}^2$.  One can solve the
dispersion relation, including the beam response, to first order in
$\chi^{\rm b}$ since its contribution is of order:
\begin{equation}
  \left(\frac{\omega_{{\rm p}{\rm b}}}{\omega_{\rm
      pe}}\right)^2\,=\,\frac{m_e}{m_p}\xi_{\rm cr}\,\ll\,1 \ .
\label{eq:oo}
\end{equation}

\subsubsection{Weibel/filamentation instability}
Consider now a mode with $k_y=0$, but $k_x\neq0$, $k_z\neq0$. The
dispersion relation, including the beam response can be written as
follows, to first order in $\chi^{\rm b}_{ij}$:
\begin{eqnarray}
  & & \left(\omega^2-\omega_{\rm
    p}^2-k^2c^2-\chi^{\rm b}_{yy}\omega^2\right)\nonumber\\ & &
  \times\Biggl[\left(\omega^2-\omega_{\rm p}^2-k_z^2c^2 +
    \chi^{\rm b}_{xx}\right)\left(\omega^2-\omega_{\rm
      p}^2-k_x^2c^2+\chi^{\rm b}_{zz}\right)\nonumber\\ & &
    \,\,\,\,-\left(k_xk_zc^2 +
    \chi^{\rm b}_{xz}\omega^2\right)^2\Biggr]\,=\,0 \ ,
\end{eqnarray}
with $\omega_{\rm p}^2\,\equiv\,\omega_{\rm pi}^2+\omega_{\rm pe}^2$. In the limit
$k_x\rightarrow 0$, one recovers the filamentation (Weibel like)
instability by developing the above dispersion relation to first order
in $\chi^{\rm b}$, with:
\begin{equation}
  \omega^2\,=\,-\omega_{{\rm p}{\rm b}}^2\frac{k^2c^2}{\omega_{\rm p}^2+k^2c^2}\ .
\end{equation}
It saturates at a growth rate $\im(\omega_{\rm  We.})\,
\simeq\,\omega_{{\rm p}{\rm b}}$ in the limit $kc\,\gg\,\omega_{\rm p}$.

\subsubsection{\v{C}erenkov resonance with oblique electrostatic modes}
In the other limit $k_z\rightarrow 0$, one can simplify the dispersion
relation for electrostatic modes down to:
\begin{equation}
  \omega^2 - \omega_{\rm p}^2 + \chi_{xx}^{\rm b}\omega^2\,\simeq\,0\ .\label{eq:long1}
\end{equation}
Then, the two stream instability resonance condition between the
Langmuir modes and the beam reads:
\begin{equation} 
  \omega\,=\,\omega_{\rm p}\left(1+\delta\right)\,=\,
\beta_{\rm b}k_x c\left(1+\delta\right)\ ,\label{eq:res1}
\end{equation}
with by assumption $\vert\delta\vert\,\ll\,1$. After insertion into
Eq.~(\ref{eq:long1}), this yields:
\begin{equation}
  \delta^3\,=\,\frac{\omega_{{\rm p}{\rm b}}^2}{2\Gamma_{\rm b}^2\omega_{\rm p}^2}\ ,
\end{equation}
hence a growth rate:
\begin{equation}
  \im(\omega)\,\simeq\,\frac{\sqrt{3}}{2^{4/3}}
  \left(\frac{\omega_{{\rm p}{\rm b}}^2\omega_{\rm
      p}}{\Gamma_{\rm b}^2}\right)^{1/3}\ .\label{eq:long}
\end{equation}
One should note that the \v{C}erenkov resonance can only take place
with plasma modes with phase velocity smaller than $c$ (refraction
index $kc/\omega(k)>1$), hence transverse modes are excluded in this
respect.

The oblique mode, with $k_z\neq0$ and a resonance as above yields a
growth rate that is larger by a factor $\Gamma_{\rm b}^{2/3}$ than
the two stream rate given in Eq.~(\ref{eq:long}) for $k_z=0$ (Bret,
Firpo \& Deutsch 2004, 2005a, b). This can be understood as
follows. The instability arises from the $xx$ component of the beam
susceptibility tensor, which dominates over the other components at
the resonance [see Eq.~(\ref{eq:chib})], and which reads:
\begin{equation}
  \chi_{xx}^{\rm b}\,=\,-\frac{\omega_{{\rm p}{\rm b}}^2}{\omega^2}
\frac{\omega^2/\Gamma_{\rm b}^2 + \beta_{\rm b}^2k_z^2c^2}
  {\left(\omega-\beta_{\rm b}k_x c\right)^2}\ .
\end{equation}
This component is suppressed by $1/\Gamma_{\rm b}^2$ when $k_z=0$,
which explains the factor appearing in the r.h.s. of
Eq.~(\ref{eq:long}). For $k_z\neq0$ however, the algebra is more
cumbersome. Nevertheless, proceeding as above, with the resonance
condition Eq.~(\ref{eq:res1}), one obtains in the limit
$\delta\,\ll\,1$ and $\beta_{\rm b}\simeq 1$:
\begin{equation}
  \delta^3 \,\simeq\, \frac{\omega_{p{\rm b}}^2}{\omega_{\rm p}^2}\frac{k_z^2}{2k^2}\ .
\end{equation}
In the limit $k_z\,\gg\,k_x\,\simeq\,\omega_{\rm p}/c$, one recovers the
growth rate of the oblique mode:
\begin{equation}
  \im(\omega)\,\simeq\,\frac{\sqrt{3}}{2^{4/3}}
\left(\omega_{{\rm p}{\rm b}}^2\omega_{\rm p}\right)^{1/3}\ .\label{eq:obl}
\end{equation}
This mode obviously grows faster than the previous two.

Obviously, the mode is quasi-longitudinal, since resonance takes place
with the electrostatic modes. However it also comprises a small
electromagnetic component, $\left\vert B_y\right\vert/\left\vert
  E_z\right\vert\,\approx\,2\left\vert\delta\right\vert$, as can be
seen by solving for the eigenmode, using the full dispersion relation
including the beam contribution.

\section{Instabilities in the presence of a mean field}

As before, we look for an instability of the upstream plasma waves,
triggered by the beam of accelerated (and shock reflected)
particles. At non-relativistic shocks, one usually considers an
interaction at the Larmor resonance. However this cannot be relevant
in the ultra-relativistic case, because the interaction must develop
on a distance scale $\lesssim \ell_{\rm F}$ which is itself much
shorter than the Larmor radius. 
The particular case of a relativistic
parallel shock wave will be briefly discussed thereafter.  Note
finally that for the frequently valid condition $\beta_{\rm A}
\Gamma_{\rm sh} \sin \theta_B \ll 1$, the precursor has a length much
larger than the minimum scale for MHD description ($\ell_{\rm MHD}/
\ell_{\rm F\vert\rm u} = \beta_{\rm A} \Gamma_{\rm sh} \sin \theta_B $),
which justifies the resonance between the beam and the MHD modes.

\subsection{Oblique magnetic field}
In order to excite fast waves of frequency higher than the Larmor
frequency, we consider again the \v{C}erenkov resonance between the
non-magnetized beam and the magnetized plasma waves: $\omega -
\mathbf{k \cdot v}_{\rm b} = 0$. Let us recall that for a
ultra-relativistic beam, the velocity distribution is strongly peaked
at $v_{\rm b}\sim c$, even if the dispersion in Lorentz factor of the beam
is significant.  We also discuss the possibility of generating the
magnetic field through a (non-resonant) Weibel (filamentation)
instability with $k_x=0$.

\subsubsection{Weibel -- filamentation instability}
This instability taking place in the shock transition layer between
the unshocked plasma and the shocked plasma has been discussed in
detail in the waterbag approximation for an unmagnetized plasma
(Medvedev \& Loeb 1999; Wiersma \& Achterberg 2004; Lyubarsky \&
Eichler 2006; Achterberg \& Wiersma 2007, Achterberg, Wiersma \&
Norman 2007). As we now argue, the Weibel instability can also proceed
in the regime of unmagnetized proton -- magnetized plasma electrons at
smaller frequencies, corresponding to the range $\omega_{\rm
  ci}\,\ll\,\omega\,\ll\,\omega_{\rm ce}$ (see also Achterberg \&
Wiersma 2007). Again, we should stress that we consider a pure ion
beam (reflected and accelerated particles), whereas most above studies
consider two neutral interpenetrating plasmas.

To simplify the algebra, we write down the dispersion relation in a
frame in which the $(\mathbf{x},\mathbf{z})$ plane has been rotated in
such a way as to align $\mathbf B$ with the third axis, denoted
$\mathbf{z}_B$; $\mathbf y$ remains the second axis $\mathbf{y}_B$. To
simplify further a cumbersome algebra, we consider a wavenumber
$\mathbf{k}\parallel\mathbf{y}_B$, perpendicular to both the beam
motion and the magnetic field. The plasma di-electric tensor is
written in this $B$ frame as:
\begin{eqnarray}
  \Lambda_{ij\vert B} \,=\,\left(\begin{array}{ccc}
    \varepsilon_1-\eta^2 & i\varepsilon_2 & 0 \\ -i\varepsilon_2 &
    \varepsilon_1 & 0 \\0 & 0 &
    \varepsilon_{\parallel}-\eta^2\end{array}\right)\ ,\label{eq:plasmadiel}
\end{eqnarray}
with the following usual definitions (for $\omega_{\rm
  ci}\,\ll\omega\,\ll\omega_{\rm ce}$):
\begin{equation}
  \varepsilon_1\,\simeq\,1 - \frac{\omega_{\rm
      pi}^2}{\omega^2}+\frac{\omega_{\rm pe}^2}{\omega_{\rm
      ce}^2}\ ,\,\,\,\, \varepsilon_2\,\simeq\, \frac{\omega_{\rm
      pe}^2}{\omega\omega_{\rm ce}}\ ,\,\,\,\,
  \varepsilon_\parallel\,\simeq\,1-\frac{\omega_{\rm
      p}^2}{\omega^2}\ .\label{eq:dielcomp-int}
\end{equation}
and $\eta\,\equiv\,kc/\omega$.  One needs to rotate the beam
susceptibility tensor to this $B$ frame. The quantity of interest will
turn out to be the $3-3$ component
$\chi^{\rm b}_{z_Bz_B}\,=\,\cos^2\theta_B\chi^{\rm b}_{xx} +
\sin^2\theta_B\chi^{\rm b}_{zz}$. To first order in $\chi^{\rm b}$, the
dispersion relation indeed has the solution:
\begin{equation}
\varepsilon_\parallel - \eta^2 + \cos^2\theta_B\chi^{\rm b}_{xx} + 
\sin^2\theta_B\chi^{\rm b}_{zz}\,=\,0 \ .
\end{equation}
Given the dependence of $\chi^{\rm b}_{xx}$ on $\omega$, this is a quartic
equation which admits the solution leading to the Weibel
(filamentation) instability:
\begin{equation}
\omega^2\,\simeq\,-\omega_{{\rm p}{\rm b}}^2\cos^2\theta_B
\frac{k^2c^2}{\omega_{\rm p}^2 + k^2c^2}\ .\label{eq:wb}
\end{equation}
As in the unmagnetized case, it saturates at a growth rate $\simeq
\omega_{{\rm p}{\rm b}}\cos\theta_B$ (up to the angular dependence on $B$). Note
that in the limit $\cos\theta_B\rightarrow 0$, this instability does
not disappear. In order to see this, one has to consider the other
branch of the dispersion relation, for $\cos\theta_B=0$,
$\mathbf{k}=k_z\mathbf{z}$:
\begin{equation}
\left(\epsilon_1-\eta^2 + \chi_{xx}^{\rm b}\right)
\left(\epsilon_1-\eta^2+ \chi_{yy}^{\rm b}\right) -\epsilon_2^2\,=\,0\ .
\end{equation}
One of the roots corresponds to the Whistler mode and the other to the
Weibel unstable mode with $\omega^2\,\simeq\,-\omega_{{\rm p}{\rm b}}^2$.

The above thus shows that fast waves can be excited by the
relativistic stream in the intermediate range between MHD and electron
dynamics, i.e. with unmagnetized plasma ions but magnetized
electrons. The typical length scale of these waves for which maximal
growth occurs is obviously the electron inertial scale $\delta_{\rm e}
\equiv c/\omega_{\rm p}$ as before.

\subsubsection{\v{C}erenkov resonance with longitudinal modes}
The previous discussion of the \v{C}erenkov instability with
electrostatic modes can be generalized to the magnetized plasma limit
by considering those modes with $\mathbf{k}\parallel\mathbf{B}$, which
do not feel the magnetic field (see Lyubarsky 2002 for a discussion of
this instability in the case of pulsar magnetospheres). Rewriting the
above plasma di-electric tensor for a wavenumber parallel to the
magnetic field, it is straightforward to see that the dispersion
relation admits the longitudinal branch given by:
\begin{equation}
\varepsilon_\parallel + \cos^2\theta_B\chi^{\rm b}_{xx} +
\sin^2\theta_B\chi^{\rm b}_{zz}\,=\,0\ .
\end{equation}
In order to avoid confusion, it may be useful to stress that the
previous {\em oblique} denomination refers to the angle between the
wavenumber and the beam direction, while the present term {\em
  longitudinal} here refers to the parallel nature of $\mathbf{k}$ and
$\mathbf{B}$.  At the \v{C}erenkov resonance $\omega=\omega_{\rm
  p}(1+\delta)=\beta_{\rm b}k_xc(1+\delta)$, with
$\vert\delta\vert\ll1$, one has $\vert\chi^{\rm
  b}_{xx}\vert\,\gg\,\vert\chi^{\rm b}_{zz}\vert$, therefore one can
obtain the following approximate solution for the growth rate:
\begin{equation}
\im(\omega)\,\simeq\,\frac{\sqrt{3}}{2^{4/3}} \left[\omega_{{\rm
      p}{\rm b}}^2\omega_{\rm p}\cos^2\theta_B\left(\frac{1}{\Gamma_{\rm b}^2} +
  \frac{k^2c^2\sin^2\theta_B}{\omega_{\rm
      p}^2}\right)\right]^{1/3}\ .\label{eq:Blong}
\end{equation}
Recalling that $k_x=k\cos\theta_B=\omega_p/c$, and that $\Gamma_{\rm
  b}^2=\Gamma_{\rm sh}^4$, one can neglect in all generality the first
term in the parenthesis, so that
$\im(\omega)\,\simeq\,\left(\omega_{{\rm p}{\rm b}}^2\omega_{\rm
  p}\sin^2\theta_B\right)^{1/3}$.

\subsubsection{Resonant instability with Alfv\'en modes}

Turning now to resonant instabilities with Alfv\'en waves, we consider
a wavector in the $(\mathbf{x},\mathbf{z})$ plane. The resonance
condition for Alfv\'en modes reads: $\beta_{\rm b}k_x \,\simeq\,\beta_{\rm
  A}k \cos\theta_k$, where $\theta_k$ represents the angle between the
wavenumber and the magnetic field direction. Since $\beta_{\rm
  A}\,<\,1$, this implies $k_x\,\ll\,k$, therefore the wavenumber is
mostly aligned along $\mathbf{z}$ and $\theta_k\,\simeq\,\pi/2 -
\theta_B$.

The plasma dielectric tensor now reads (we omitted negligible
contributions in $\sin^2 \theta_k$):
\begin{eqnarray}
  \Lambda_{ij\vert B}
  \,=\,\left(\begin{array}{ccc}
\varepsilon_1-\eta^2\cos^2\theta_k  &  i\varepsilon_2 &
\eta\cos\theta_k\sin\theta_k \\
-i\varepsilon_2 & \varepsilon_1 -\eta^2& 0 \\
\eta\cos\theta_k\sin\theta_k & 0 & 
\varepsilon_{\parallel}-\eta^2\sin^2\theta_k\end{array}\right)\ ,
\label{eq:diel-B}
\end{eqnarray}
with ($\omega\,\ll\,\omega_{\rm ci}$):
\begin{equation}
\varepsilon_1\,\simeq\,\frac{1}{\beta_A^2} \ ,\,\, 
\varepsilon_2\,\simeq\,0\ ,\,\,
\varepsilon_\parallel\,\simeq\, - \frac{\omega_{\rm p}^2}{\omega^2}\ .
\end{equation}
The beam susceptibility can be approximated accurately by neglecting
all components in front of $\chi^{\rm b}_{xx}$, which dominates at the
resonance, as explained above. The relevant components then are:
\begin{eqnarray}
  \chi^{\rm b}_{x_Bx_B}&\,\simeq\,& \sin^2\theta_B\chi^{\rm b}_{xx}\ ,\,\,
  \chi^{\rm b}_{z_Bz_B}\,\simeq\, \cos^2\theta_B\chi^{\rm b}_{xx}\ ,\nonumber\\
  \chi^{\rm b}_{x_Bz_B}&\,=\,&\chi^{\rm b}_{z_Bx_B}\,\simeq\, 
  \sin\theta_B\cos\theta_B\chi^{\rm b}_{xx}\  .\label{eq:chib-app}
\end{eqnarray}

The dispersion relation then takes the form:
\begin{eqnarray}
  &&\left(\frac{\omega^2}{\beta_A^2} - k^2c^2\cos^2\theta_k\right)
  \left(\omega_{\rm p}^2 + k^2c^2\sin^2\theta_k\right)\nonumber \\
  && + k^4c^4\sin^2\theta_k\cos^2\theta_k - \omega^4A_{xx}\chi^{\rm b}_{xx}\,=\,0\ ,
\end{eqnarray}
where $A_{xx}\,\simeq\, -\sin^2\theta_B\omega_{\rm p}^2/\omega^2$ in the
limit $k \delta_{\rm e} \,\ll\,1$. Writing down the resonance condition
$\omega\,=\,\beta_A k\cos\theta_k
c\left(1+\delta\right)\,=\,\beta_{\rm b}k_xc\left(1+\delta\right)$,
with $\vert\delta\vert\,\ll\,1$ as before, one obtains the growth
rate:
\begin{equation}
\im(\omega)\,\simeq\,\frac{\sqrt{3}}{2^{4/3}}\,\left(\omega_{{\rm p}{\rm b}}^2 
\beta_{\rm A}kc\cos\theta_k\right)^{1/3}\ ,
\end{equation}
where we approximated $k_z\,\simeq\,k$; recall furthermore that
$\cos\theta_k\,\simeq\,\sin\theta_B$. This instability disappears in
the limit of a parallel shock wave as one can no longer satisfy the
\v{C}erenkov resonance condition.

One should stress that the above perturbative treatment remains valid
as long as the condition $\vert\delta\vert\,\ll\,1$, which amounts to
$\xi_{\rm cr}\,\ll\,\beta_{\rm A}^2$ at maximum wave growth rate
($kc=\omega_{\rm ci}$). Therefore Alfv\'en growth is limited to
strongly magnetized shock waves only.

In the continuity of right Alfv\'en waves (the left modes being
absorbed at the ion-cyclotron resonance), there are Whistler waves for
quasi parallel propagation (with respect to the mean field), that are
electromagnetic waves with a dominant magnetic component. For quasi
perpendicular propagation, there are the ionic extraordinary modes,
which have frequencies between the ion-cyclotron frequency and the
low-hybrid frequency (obtained for large refraction index) and which
are mostly electrostatic with a weaker electromagnetic component.  For
scattering purpose, the whistler waves are the most interesting in
this intermediate range; they are actually excited in the foot of
non-relativistic collisionless shocks in space plasmas. But for
pre-heating purposes, the extraordinary ionic modes are more
interesting (they are actually used for additional heating in
tokamaks). Let us now discuss these in turn.

\subsubsection{Resonant instability with Whistler waves}\label{sec:whistler}
We proceed as before, using the plasma di-electric tensor
Eq.~(\ref{eq:diel-B}) in the range
$\omega_{\rm ci}\,\ll\,\omega\,\ll\,\omega_{\rm ce}$ with the components given
in Eq.~(\ref{eq:dielcomp-int}).  The Whistler branch of the dispersion
relation reads, to first order in the beam response $\chi^{\rm b}$
approximated by Eq.~(\ref{eq:chib-app}):
\begin{equation}
\left(\epsilon_1-\eta^2\cos^2\theta_k+\chi^{\rm
    b}_{xx}\sin^2\theta_B\right)
\left(\epsilon_1 - \eta^2\right)
-\epsilon_2^2\,=\,0\ .
\end{equation}
When the beam response is absent, one recovers the dispersion relation
for oblique Whistler waves:
\begin{equation}
\omega_{\rm Wh.}^2\,\simeq\,\frac{\omega_{\rm ce}^2}
{\omega_{\rm pe}^4}k^4c^4\cos^2\theta_k\ .
\end{equation}
Introducing the resonance $\omega=\omega_{\rm
  Wh.}\left(1+\delta\right)\,=\,\beta_{\rm b}k_x
c\left(1+\delta\right)$, with $\vert\delta\vert\,\ll\,1$, we obtain
the growth rate:
\begin{equation}
\im(\omega)\,\simeq\,\frac{\sqrt{3}}{2^{4/3}}
\left(\omega_{{\rm p}{\rm b}}^2\omega_{\rm Wh.}\right)^{1/3}\ .
\end{equation}
In the latter equation, we again approximated $k_z\,\simeq\,k$, since
the resonance condition implies $k_x\,\ll\,k$ (therefore
$\cos\theta_k\,\simeq\,\sin\theta_B$). The instability disappears in
the limit of a parallel shock wave as well, because the resonance
condition cannot be satisfied. Maximum growth occurs here as well for
$k\,\simeq\,c/\omega_{\rm pe}\,\simeq\,c/\omega_{\rm p}$, i.e. at the electron
inertial scale $\delta_{\rm e}$, however the excitation range extends to the
proton inertial scale $\delta_{\rm i}$ where it matches with the Alfv\'en
wave instability.

As before, the perturbative treatment remains valid as long as
$\vert\delta\vert\,\ll\,1$, which amounts to $\xi_{\rm
  cr}\,\ll\,(m_p/m_e)^2\beta_{\rm A}^2$. This condition is more easily
satisfied that the corresponding one for amplification of Alfv\'en
waves. It will be discussed in more detail in
Section~\ref{sec:disclim}.

\subsubsection{Resonant instability with extraordinary modes}

At MHD scales, the extraordinary ionic modes (that propagate with wave
vectors almost perpendicular to the magnetic field) assimilate to
magneto-sonic modes. These modes has been shown to be unstable when
there is a net electric charge carried by the cosmic rays (Pelletier,
Lemoine \& Marcowith 2009). The obtained growth rates are increasing
with wave numbers indicating an instability that reaches its maximum
growth at scales shorter than the MHD range. Let us therefore discuss
how this instability extends to sub-MHD scales.

Let us first discuss the ionic (lower hybrid) branch,
$\omega\,<\,\omega_{\rm lh}$, with $\omega_{\rm
  lh}\equiv\sqrt{\omega_{\rm ci}\omega_{\rm ce}}$. In the following,
we assume for simplicity $\omega_{\rm ce}\,\ll\,\omega_{\rm pe}$,
i.e. a weakly magnetized plasma. In the $B$ frame, in which
$\mathbf{B}$ is along $\mathbf{z}_B$ and the beam propagates in the
$(\mathbf{x},\mathbf{z})$ plane, take
$\mathbf{k}\parallel\mathbf{y}_B$, with a small component $k_{x_B}$,
i.e. in the $(x,z)$ plane but perpendicular to $B$. The dispersion
relation to zeroth order in $\chi^{\rm b}$ reads:
\begin{equation}
\eta^2\,=\,\frac{\epsilon_1^2-\epsilon_2^2}{\epsilon_1}\ .
\end{equation}
with (since $\omega\,<\,\omega_{\rm lh}\,\ll\,\omega_{\rm ce}$):
\begin{equation}
  \frac{\epsilon_1^2-\epsilon_2^2}{\epsilon_1}\,\simeq\,
  \frac{\omega_{\rm ce}^2}{\omega_{\rm ci}^2\omega_{\rm pe}^2}
  \frac{\omega^2\omega_{\rm ci}^2-\left(\omega_{\rm ci}^2+\omega_{\rm pi}^2\right)^2}
{\omega^2-\omega_{\rm lh}^2}\ ,
\end{equation}
hence
\begin{eqnarray}
  \frac{\epsilon_1^2-\epsilon_2^2}{\epsilon_1} &\,\simeq\,& 
\frac{\omega_{\rm pi}^2}{\omega_{\rm ci}^2}\quad (\omega\,\ll\,\omega_{\rm ci})\ ,
  \nonumber\\
  \frac{\epsilon_1^2-\epsilon_2^2}{\epsilon_1} &\,\simeq\,& 
\frac{\omega_{\rm pe}^2}{\omega_{\rm lh}^2-\omega^2}
\quad(\omega_{\rm ci}\,\ll\omega\,\ll\omega_{\rm lh})\ .
\end{eqnarray}
At $\omega\,\ll\omega_{\rm ci}$, this gives the fast magnetosonic
branch with $\omega_{\rm H}\,\simeq\,\beta_{\rm A}kc$, while at
$\omega_{\rm ci}\,\ll\omega\,\ll\,\omega_{\rm lh}$, $\omega_{\rm
  H}\,\sim\,\omega_{\rm lh}kc/\sqrt{k^2c^2+\omega_{\rm pe}^2}$.  We
define:
\begin{equation}
{\cal D}(k,\omega)\,\equiv\,\frac{\epsilon_1^2-\epsilon_2^2}{\epsilon_1} - \eta^2\ .
\end{equation}
so that:
\begin{eqnarray}
  \omega^2\frac{\partial}{\partial \omega^2}{\cal
    D}(k,\omega)\,\simeq\, \eta^2 
\quad (\omega\,\ll\,\omega_{\rm ci})\ ,
  \nonumber\\
  \omega^2\frac{\partial}{\partial \omega^2}{\cal
    D}(k,\omega)\,\simeq\, 
\eta^2 \frac{\omega_{\rm lh}^2}{\omega_{\rm lh}^2-\omega^2} \quad 
(\omega_{\rm ci}\,\ll\omega\,\ll\omega_{\rm lh})\ .
\end{eqnarray}
Including the beam response, the dispersion relation becomes:
\begin{equation}
\epsilon_1^2 - \epsilon_2^2 - \epsilon_1\eta^2 + 
\left(\epsilon_1-\eta_{x_B}^2\right)\sin^2\theta_B\chi^{\rm b}_{xx}\,=\,0\ .
\end{equation}
We neglect the term $\eta_{x_B}^2\,\ll \eta^2$ in front of
$\epsilon_1\,\sim\,1/\beta_{\rm A}^2$ (at
$\omega\,\ll\,\omega_{\rm ci}$). At the resonance $\omega\,=\,\omega_{\rm
  H}(1+\delta)$, with $\omega_{\rm H}$ the solution of ${\cal
  D}(k,\omega_{\rm H})\,=\,0$, one finds:
\begin{equation}
  \delta^3\,\simeq\,\frac{1}{2}\frac{\omega_{{\rm
        p}{\rm b}}^2\sin^2\theta_B}{\omega_{\rm H}^2}
  \left[\omega^2\frac{\partial}{\partial\omega^2}{\cal
      D}(k,\omega)\right]^{-1} \frac{k_y^2c^2}{\omega_{\rm
      H}^2}\ . \label{eq:gH}
\end{equation}
The growth rate for \v{C}erenkov resonance with the lower hybrid
extraordinary mode thus reads:
\begin{eqnarray}
\im(\omega_{\rm LX})&\,\simeq\,&\frac{\sqrt{3}}{2^{4/3}}
\left(\omega_{{\rm p}{\rm b}}^2\sin^2\theta_B\frac{k_y^2}{k^2}\beta_{\rm
  A}kc\right)^{1/3} \quad (\omega\,\ll\,\omega_{\rm
  ci})\ ,\nonumber\\ \im(\omega_{\rm
  LX})&\,\simeq\,&\frac{\sqrt{3}}{2^{4/3}} \left[\omega_{{\rm
      p}{\rm b}}^2\sin^2\theta_B\frac{k_y^2}{k^2}\frac{\omega_{\rm
      lh}\omega_{\rm pe}^2kc} {\left(k^2c^2+\omega_{\rm
      pe}^2\right)^{3/2}}\right]^{1/3} \nonumber\\ & &
\quad\quad\quad\quad (\omega_{\rm ci}\,\ll\,\omega\,\ll\,\omega_{\rm
  lh})\ .
\end{eqnarray}
In the limit of magnetosonic modes, $\omega \, \ll \, \omega_{\rm
  ci}$, one recovers the same growth rate as for Alfv\'en waves; note
that $\beta_{\rm A}kc\,\ll\,\omega_{\rm ci}$ implies
$k\,\ll\,\omega_{\rm pi}/c$. At smaller scales, one finds that the
growth rate reaches its maximum at $k\,\simeq\,\omega_{\rm pe}/c$ with
$\im(\omega_{\rm LX})\,\sim\,(\omega_{{\rm
    p}{\rm b}}^2\sin^2\theta_B\omega_{\rm lh})^{1/3}$. We can expect this
instability to provide efficient heating of the protons in the foot.

Turning to the electronic (upper hybrid) modes, around
$\omega\,\sim\,\omega_{\rm pe}$, one obtains:
\begin{eqnarray}
  \frac{\epsilon_1^2-\epsilon_2^2}{\epsilon_1}&\,\simeq\,&
\frac{(\omega^2-\omega_x^2)(\omega^2-\omega_z^2)}
  {\omega_{\rm pe}^2(\omega^2-\omega_{\rm uh}^2)}\  ,
\end{eqnarray}
with $\omega_x\,\simeq\,\omega_{\rm pe}-\omega_{\rm ce}/2$,
$\omega_z\,\simeq\,\omega_{\rm pe}+\omega_{\rm ce}/2$ and
$\omega_{\rm uh}\,\equiv\,\left(\omega_{p}^2+\omega_{\rm ce}^2\right)^{1/2}$.
The dispersion relation takes the same form ${\cal D}(k,\omega)=0$, but now:
\begin{equation}
\frac{\partial}{\partial\omega^2}{\cal D}(k,\omega)\,\simeq\,
\eta^2\left(\frac{\omega^2}{\omega^2-\omega_x^2}+
\frac{\omega^2}{\omega^2-\omega_z^2} -
\frac{\omega^2}{\omega^2-\omega_{\rm uh}^2}+1 \right)\ .
\end{equation}
The growth rate can be written in the same algebraic form as
(\ref{eq:gH}). It vanishes in both limits $\omega\rightarrow\omega_x$
and $\omega\rightarrow\omega_z$, while for
$\omega\,\simeq\,\omega_{\rm pe}$, giving $\eta\,\simeq\,1$, one
obtains:
\begin{equation}
  \im(\omega_{\rm
    UX})\,\simeq\,\frac{\sqrt{3}}{2^{4/3}}\left(\omega_{{\rm
      p}{\rm b}}^2\sin^2\theta_B \omega_{\rm pe}\frac{\omega_{\rm
      ce}^2}{\omega_{\rm pe}^2}\frac{k_y^2}{k^2}\right)^{1/3}\ .
\end{equation}
It vanishes in the limit $\omega_{\rm ce}/\omega_{\rm
  pe}\rightarrow0$, in which limit the electronic extraordinary branch
actually disappears.

Being electrostatic in nature, these waves participate mostly to the
heating process in the shock foot or precursor. However their
scattering efficiency is comparable to the magnetic perturbations as
will be seen further on.

\subsection{The particular case of a parallel magnetic field}\label{sec:bell}

When the magnetic field is almost parallel, i.e. $\theta_B < 1/
\Gamma_{\rm sh}$, the relativistic Bell non-resonant instability (Bell
2004, 2005) can develop (e.g. Milosavljevi\'c \& Nakar 2006; Reville,
Kirk \& Duffy 2006). This instability is triggered by the charge
current carried by the cosmic rays in the precursor, which induces a
return current in the plasma, thereby destabilizing non-resonant waves
of wavelength shorter than the typical Larmor radius, the cosmic rays
being unresponsive to the excitation of the waves. The growth rate of
this instability in the upstream frame is (Reville, Kirk \& Duffy
2006):
\begin{equation}
\im\left(\omega_{\rm Bell}\right)\,\simeq\,
\frac{\beta_{\rm b}n_{{\rm b}\vert\rm u}}{n_{\rm u}}\omega_{\rm pi}\ ,
\end{equation}
and growth is maximal at the scale $k_c\,\simeq\,\im(\omega_{\rm
  Bell})/(\beta_{\rm A}c)$.

One can then verify that, under quite general assumptions, this growth
rate is larger than the growth rate of the Weibel instability, since
the ratio of these two is given by:
\begin{equation}
  \frac{\im\left(\omega_{\rm Bell}\right)}{\im\left(\omega_{\rm
        We.}\right)}
  \,\simeq\,\Gamma_{\rm sh}^2\xi_{\rm cr}^{1/2}\ .
\end{equation}
One must emphasize however that the Bell instability is quenched when
the growth rate exceeds the ion cyclotron frequency see Couch,
Milosavljevic \& Nakar (2008), Riquelme \& Spitkovsky (2009), Ohira et
al. (2009).  This limitation will be made clear in
Section~\ref{sec:parshock}.

\section{Limitations of the instabilities}\label{sec:disclim}

Using the growth rates derived previously, we can now delimit the
conditions under which the various instabilities become effective, and
which one dominates. We then discuss the limit between unmagnetized
and magnetized shock waves, from the point of view of these upstream
instabilities. 

\subsection{Superluminal shock waves}

In this Section, we discuss the generic case of relativistic
superluminal shock waves, taking $\sin^2\theta_B\,\sim\,1$. Unless
otherwise noted, we assume an $e-p$ plasma; we will discuss how the
results are modified in the limit of a pair plasma at the end of this
discussion. The more particular case of relativistic parallel shock
waves is treated further below.

We start by introducing the two parameters $X$ and $Y$ defined as
follows:
\begin{eqnarray}
  X&\,\equiv\,& \Gamma_{\rm sh}\frac{m_e}{m_p}\ ,\nonumber\\
  Y &\,\equiv\,& \Gamma_{\rm sh}^4\frac{B_{\rm u\vert\rm u}^2}
{4\pi n_{{\rm b}\vert\rm u}m_pc^2}\,=\,\Gamma_{\rm sh}^2\sigma_{\rm u}\xi_{\rm cr}^{-1}\ .
\end{eqnarray}
The upstream magnetization parameter $\sigma_{\rm u}$ also corresponds
to the Alfv\'en velocity squared of the upstream plasma (in units of
$c^2$). If the field is fully perpendicular, the shock crossing
conditions imply $B_{\rm d\vert d,\perp}\,\simeq\, B_{\rm u\vert\rm
  u,\perp}\Gamma_{\rm sh}\sqrt{8}$, and for the enthalpy $h_{\rm
  d\vert d}\,\simeq\, (8/3)\Gamma_{\rm sh}^2h_{\rm u\vert\rm u}$ (for
a cold upstream plasma, see Blandford \& McKee 1976), so that
$\sigma_{\rm d}\,\simeq\,3\sigma_{\rm u}\sin^2\theta_B$. If the
magnetic field is mostly parallel, meaning
$\sin\theta_B\,\leq\,1/\Gamma_{\rm sh}$, then $\sigma_{\rm
  d}\,\sim\,(3/8)\Gamma_{\rm sh}^{-2}\sigma_{\rm u}$.

Let us first compare the growth rates of the instabilities obtained in
the magnetized case; the unmagnetized case (in particular the oblique
mode) will be discussed thereafter. We carry out this comparison at
the wavenumber where the growth rates reach their maximum, namely $k
\,\sim\,\omega_{\rm pe}/c$. The ratio of the Weibel to Whistler
instability growth rates is given by:
\begin{equation}
\frac{\im\left(\omega_{\rm We.}\right)}{\im\left(\omega_{\rm Wh.}\right)}\,=\,
\left(\frac{X}{Y}\right)^{1/6}\ ,
\end{equation}
hence the Weibel instability will dominate over the Whistler
\v{C}erenkov resonant instability whenever $Y\,\ll\,X$. The fastest
mode however corresponds to the \v{C}erenkov resonance with the
longitudinal modes along the magnetic field, since the ratio of the
growth rates of this mode to the Weibel mode is
$(m_p/m_e)^{1/6}\xi_{\rm cr}^{-1/6}$, which is always greater than
one.

Since the \v{C}erenkov resonant instabilities for the Whistler and
Alfv\'en waves scale in a similar way with the eigenfrequencies of the
resonant plasma modes, it is straightforward to see that Whistler
waves will always grow faster than the Alfv\'en waves.

Concerning the extraordinary modes, one finds that $\im(\omega_{\rm
  Wh.})/\im(\omega_{\rm LX})\,\sim\,(m_p/m_e)^{1/6}$ on the ionic
(lower hybrid) branch, while $\im(\omega_{\rm Wh.})/\im(\omega_{\rm
  UX})\,\sim\,(\omega_{\rm pe}/\omega_{\rm ce})^{1/3}$ on the
electronic (upper hybrid) branch. Therefore the growth of these modes
is always sub-dominant with respect to that of Whistler and Weibel
modes. Since the growth rates of the Alfv\'en and extraordinary modes
are always smaller than that of the Whistler modes, we discard the
former in the following.

Additional constraints can be obtained as follows. First of all, the
above derivation of the instabilities has assumed the beam to be
unmagnetized, i.e. that the growth time be much shorter than the
Larmor time of the beam particles. This condition is always easily
satisfied, since it reads: $Y\,\ll\,\Gamma_{\rm sh}^6$ for the Weibel
instability, $Y\,\ll\,\Gamma_{\rm sh}^6\xi_{\rm
  cr}^{-1/3}(m_e/m_p)^{-1/3}$ for the longitudinal mode and
$Y\,\ll\,\Gamma_{\rm sh}^8m_p/m_e$ for the Whistler \v{C}erenkov
resonant mode. One can explicit the dependence of $Y$ on the shock
parameters in order to verify this; for the Weibel instability, the
condition amounts to $\xi_{\rm cr}\,\gg\,\Gamma_{\rm
  sh}^{-4}\sigma_{\rm u}$, which is indeed easily verified at large
Lorentz factors.

Concerning the \v{C}erenkov resonant modes, one must also require that
$\vert\delta\vert\,<\,1$ in order for the perturbative treatment to be
apply. In the case of longitudinal modes, this is automatically
satisfied since $\omega_{{\rm p}{\rm b}}<\omega_{\rm p}$. However, for
Whistler modes of smaller eigenfrequency, this implies a non-trivial
constraint $\omega_{{\rm p}{\rm b}}<\omega_{\rm Wh.}$ which can be
translated into $Y\,\gg\,X^2$ for $\omega_{\rm Wh.}=\omega_{\rm ce}$.

Further bounds can be obtained by requiring that the background
protons are non-magnetized in the case of the Weibel instability,
which requires $\im\left(\omega\right)\,\gg\,\omega_{\rm ci}$. This
condition is however superseded by the requirement that the growth can
occur on the precursor length scale, since $\ell_{\rm F}/c\sim
(\Gamma_{\rm sh}\omega_{\rm ci})^{-1}$ [see Eq.~(\ref{eq:lfb})].  At
this stage, it is important to point out a fundamental difference
between the \v{C}erenkov resonant instabilities and the Weibel /
filamentation instabilities. The former have, by definition of the
resonance, a phase velocity along the shock normal which, to zeroth
order in $\vert\delta\vert$ exceeds the shock velocity, while the
latter have vanishing phase velocity along $\mathbf{x}$. Therefore the
timescale available for the growth of these non-resonant waves is the
crossing time of the precursor: they are sourced at a typical distance
$\ell_{\rm F}$ away from the shock, then advected downstream on this
timescale. Regarding the resonant modes, their phase velocity along
$\mathbf{x}$ is $\beta_{\phi,x}=\beta_{\rm b}(1 + \delta_R)$, with
$\delta_R={\cal R}(\delta)$. Since $\delta_{\rm R}<0$ for the resonant
modes, one must consider three possible cases: (i)
$\beta_{\phi,x}<\beta_{\rm sh}$, in which case the mode is advected
away on a timescale $\ell_{\rm F}/c$ as for the non-resonant modes;
(ii) $\beta_{\phi,x}>\beta_{\rm sh}$, in which case the mode
propagates forward, but exits the precursor (where it is sourced) on a
similar timescale; and (iii) $\beta_{\phi,x}\,\simeq\,\beta_{\rm sh}$,
in which case the mode can be excited on a timescale
$\simeq\,c^{-1}\ell_{\rm F}/(\beta_{\rm sh}-\beta_{\phi,x})$ and where
the divergence corresponds to the situation of a mode surfing on the
shock precursor. However condition (i) appears to be the most likely,
as least in the ultra-relativistic limit, for it amounts to
$2\Gamma_{\rm sh}^2\vert\delta_{\rm R}\vert\,\gg\,1$. Indeed, all
resonant instabilities have a growth rate $\sim (\omega_{\rm p
  {\rm b}}^2\omega)^{1/3}$ where $\omega$ is the eigenfrequency of the
resonant mode (an exception is the upper hybrid mode for which the
growth rate is smaller by $(\omega_{\rm ce}/\omega_{\rm pe})^{2/3}$,
in which case the following condition is even stronger), therefore the
condition $2\Gamma_{\rm sh}^2\vert\delta_{\rm R}\vert\,\gg\,1$ can be
rewritten as $\omega/\omega_{\rm pe}\,\ll\,7 (\Gamma_{\rm
  sh}/10)^3(\xi_{\rm cr}/0.1)^{1/2}$, which is generically
satisfied. This means that the phase velocity of the resonant modes,
when corrected by the effect of the beam becomes smaller than the
shock front velocity, so that these modes are advected on a timescale
$\sim\ell_{\rm F}/c$ and transmitted downstream, after all. For the
purpose of magnetic field amplification downstream and particle
acceleration, this is certainly noteworthy, as such true plasma
eigenmodes (Whistler, Alfv\'en, extraordinary and electrostatic
longitudinal or oblique modes) can be expected to have a longer
lifetime than the Weibel modes.

The modes thus grow on the precursor crossing timescale if
$\im\left(\omega\right)\ell_{\rm F}/c\,\gg\,1$, which can be recast as
$Y\,\ll\,1$ for the Weibel instability, $Y\,\ll\,\xi_{rm
  cr}^{-1/3}(m_e/m_p)^{-1/3}$ and $XY\,\ll\,1$ for the \v{C}erenkov
resonant Whistler mode. Henceforth, we use the parameter
$G\,\equiv\,(\omega_{\rm p{\rm b}}/\omega_{\rm p})^{-2/3}=\xi_{\rm
  cr}^{-1/3}(m_e/m_p)^{-1/3}\,>\,1$.

In short, we find that the various instabilities discussed here are
more likely quenched by advection rather than by saturation. In
Section~\ref{sec:appl}, we provide several concrete estimates for
cases of astrophysical interest and it will be found that this limit
is indeed quite stringent.

Finally, one must also require that the growth rate of the
\v{C}erenkov resonant instabilities does not exceed the proper
eigenfrequency of the mode. For the Whistler modes, as discussed at
the end of Section~\ref{sec:whistler}, this implies $Y\,\gg\,X^2$.

\begin{figure}
\includegraphics[width=0.5\textwidth]{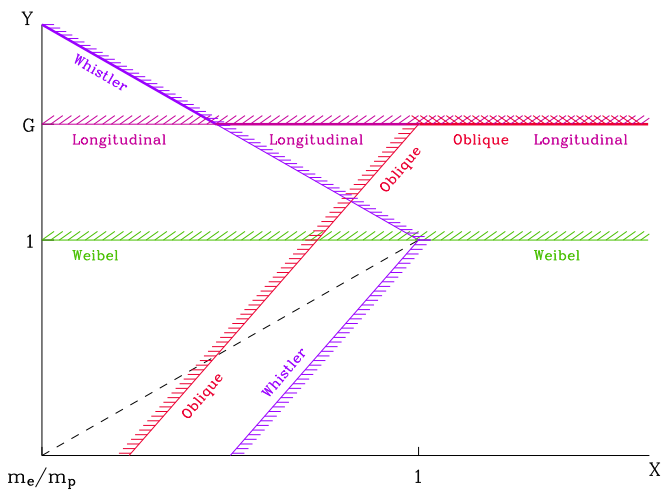}
\caption{Instability diagram for superluminal relativistic shock
    waves (assuming $\sin^2\theta_B\sim 1$): in abscissa,
    $X\,\equiv\,\Gamma_{\rm sh}m_e/m_p$, in ordinates $Y \,\equiv\,
    \Gamma_{\rm sh}^4B_{\rm u}^2/ \left(4\pi n_{{\rm b}\vert\rm
      u}m_pc^2\right)$. The parameter $G\,=\,\xi_{\rm
      cr}^{-1/3}(m_e/m_p)^{-1/3}>1$. The axes are plotted in log-log
    on arbitrary scale. The main result is summarized by the thick
    solid line, which indicates the maximum value of $Y(X)$ which
    allows electromagnetic waves to grow. The other lines indicate the
    boundaries of the regions of growth of the various instabilities,
    as indicated. The hierarchy of growth rates, from largest to
    smallest is as follows: oblique and longitudinal, then Whistler
    and/or Weibel.  The long dashed line separates the regions in
    which the growth of Whistler or Weibel modes is faster: for values
    of $Y(X)$ larger than the long dashed line, Whistler modes grow
    faster. The growth rates of the oblique mode and the longitudinal
    mode are comparable. Unlike the longitudinal mode, the oblique
    instability is limited by the assumption of unmagnetization (see
    main text), but at the same time, it applies to a larger
    wavenumber phase space.  The regions for Alfv\'en and
    extraordinary modes are not indicated (see main text).
\label{fig:XY_perp}}
\end{figure}

In Section~3, we have also examined the growth rates in the absence of
a mean magnetic field, and concluded that the oblique mode of Bret,
Firpo \& Deutsch (2004, 2005a, b) was by far the fastest. This
instability is very similar to the \v{C}erenkov resonance with the
longitudinal modes propagating along the magnetic field and indeed the
growth rates only differ by $\sin^{2/3}\theta_B$, see
Eqs.~(\ref{eq:obl}),(\ref{eq:Blong}). The difference lies in the
degree of magnetization of the ambient plasma: while the oblique mode
is limited to the unmagnetized limit, the longitudinal mode does not
suffer from such constraint; the oblique mode however covers a larger
fraction of the wavenumber phase space than the longitudinal mode.

With respect to the oblique mode instability, the shock can be
described as unmagnetized as long as the background electrons and
protons remain unmagnetized on the timescale of the instability; of
course, one must also require that the instability has time to grow on
the length scale of the precursor. Note that the latter condition also
implies that the beam can be considered as unmagnetized over the
instability growth timescale, which is another necessary
condition. For the oblique modes, those conditions amount to:
\begin{eqnarray}
  \im\left(\omega_{\rm obl.}\right)\,\gg\,\omega_{\rm ce} &
  \Leftrightarrow & Y\,\ll\,G
  X^2\label{eq:obllimmag}\ ,\\ \im\left(\omega_{\rm
    obl.}\right)\,\gg\,c/\ell_{\rm F} & \Leftrightarrow &
  Y\,\ll\,G
  \ ,\label{eq:obllimadv}
\end{eqnarray}
with $G\,=\,\xi_{\rm cr}^{-1/3}(m_e/m_p)^{-1/3}\,>\,1$ as above.
Provided the above two conditions are satisfied, the oblique mode
dominates over the Weibel and Whistler \v{C}erenkov instability growth
rates, just as the longitudinal mode.  The \v{C}erenkov resonant
instability with Whistler waves dominates over the oblique modes when
$X\,\lesssim\,G^{-1/3}$ and $G
X^{2}\,\lesssim\,Y\,\lesssim\,1/X$. The \v{C}erenkov resonant
  instability with Whistler waves dominates over the longitudinal
  modes when $X\,\lesssim\,G^{-1}$ and
  $G\,\lesssim\,Y\,\lesssim\,1/X$.  For $X\,\lesssim\,G^{-1}$ and
  $Y\,\gtrsim\,X^{-1}$, or for $G^{-1}\,\lesssim\,X$ and $Y\,\gtrsim\,
  G$ neither of the above instabilities can grow. For reference,
  $X\,\lesssim\,G^{-1}$ corresponds to $\Gamma_{\rm
    sh}\,\lesssim\,150 \xi_{\rm cr}^{1/3}$.  The above regions can be
summarized in the $X-Y$ plane as in Fig.~\ref{fig:XY_perp}, which
delimit the domains in which the various instabilities can grow, and
which of these instabilities dominates in each case.

From the above discussion, the case of a pair shock is easily obtained
by taking $m_p/m_e\,\rightarrow\,1$, by restricting oneself to the
study of the oblique and Weibel modes, and by considering only the
right part of Fig.~\ref{fig:XY_perp} with $X>1$, since $X=\Gamma_{\rm
  sh}$ for a pair shock. One sees that, irrespectively of $\Gamma_{\rm
  sh}$, the oblique and longitudinal modes can grow if
$Y\,\lesssim\,\xi_{cr}^{-1/3}$ and the Weibel mode grows on the
precursor timescale if $Y\,\lesssim\,1$.

In this respect, it is instructive to compare the present results with
the latest simulations of Sironi \& Spitkovsky (2009). These authors
find that the growth of instabilities is quenched when the
magnetization $\sigma_{\rm u}\,\gtrsim\,0.03$ for a perpendicular (or
oblique) pair shock with $\Gamma_{\rm sh}\,\simeq\,20$. This
corresponds to $X\,\simeq\,20$ and $Y\,\simeq\,10\xi_{\rm
  cr}^{-1}(\sigma_{\rm u}/0.03)$. Our results indicate that indeed, at
this high level of magnetisation, both Weibel and oblique/longitudinal
instabilities are quenched by advection. Note that these simulations
do not exclude that the instabilities are quenched even at lower
magnetisations. Our calculations thus bring to light the following
point of interest. One should not infer from the simulations of Sironi
\& Spitkovsky (2009) that superluminal shock waves cannot lead to
magnetic field amplification. This conclusion entirely depends on the
level of magnetization. It would therefore be interesting to extend
the PIC simulations down to weakly magnetized shocks with $\sigma_{\rm
  u}\,\sim\, 3\times 10^{-3}\xi_{\rm cr}^{2/3}(\Gamma_{\rm
  sh}/20)^{-2}$ in order to probe the limit at which the oblique mode
can grow.

\subsection{Parallel shock waves}\label{sec:parshock}

In the ultra-relativistic limit, parallel shock waves are non-generic;
however, they may lead more easily to particle acceleration than
superluminal shock waves (since the argument discussed in Lemoine,
Pelletier \& Revenu 2006 no longer applies) and consequently provide
interesting observational signatures. One can extend the above
discussion to the case of parallel shock waves as follows.

First of all, the main limitation of the instabilities, that is due to
the precursor crossing timescale disappears in the limit $\Gamma_{\rm
  sh}\sin\theta_B\rightarrow 0$ as a fraction 
\begin{equation}
p\,\equiv\,\frac{1-\beta_{\rm sh} \cos \theta_B - \frac{1}{\Gamma_{\rm
      sh}} \sin \theta_B}{1-\beta_{\rm sh}}\,\simeq \,(1-\Gamma_{\rm sh}
\theta_B)^2
\end{equation}
 of the particles can propagate to upstream infinity (at least in the
 limit of a fully coherent magnetic field). This can be seen as
 follows. Particles cross the shock wave back toward downstream once
 their angle cosine with the shock normal becomes smaller than
 $\beta_{\rm sh}$. However, when $\Gamma_{\rm sh}\sin\theta_B<1$,
 there exists a cone ${\cal C}_B$ around the magnetic field direction,
 of opening angle $\theta_B - {\rm acos}(\beta_{\rm sh})$, that never
 intersects the cone ${\cal C}$ defined around the shock normal with
 opening angle ${\rm acos}(\beta_{\rm sh})$ . Because the pitch angle
 of the particles with respect to the magnetic field direction is
 conserved, particles that enter toward upstream in this cone ${\cal
   C}_B$ never recross the shock toward downstream (up to the
 influence of the turbulence). The fraction of particles that enter in
 this cone is approximately given by the ratio of the solid angles,
 i.e. the factor $(1-\Gamma_{\rm sh}\theta_B)^2$ quoted
 before. Depending on the value of $\Gamma_{\rm sh}\theta_B$, this
 fraction can be substantial and these particles can excite plasma
 waves up to large distances from the shock. Of course, the actual
 precursor length remains finite as a result of the influence of large
 and short scale turbulence. In the following, we will simply discard
 these advection constraints in order to avoid introducing new
 parameters.

We also choose to discuss the limitations as a function of $X$ and
$Y$, but at a fixed value of $\Gamma_{\rm sh}\sin\theta_B\,<\,1$. The
fact that $X\,\propto\,\Gamma_{\rm sh}$ and that $\Gamma_{\rm
  sh}\sin\theta_B$ is fixed modifies slightly the limitations derived
previously. The main limitation for the oblique mode is the
non-magnetization condition, $\im(\omega_{\rm obl.})\,\gg\,\omega_{\rm
  ce}$ which can be rewritten $Y\,\ll\, G X^2$ as before. For the
Weibel mode, the condition of non-magnetization of the protons,
$\im(\omega_{\rm We.})\,\gg\,\omega_{\rm ci}$ now reads
$Y\,\ll\,(m_p/m_e)^2X^2$.

The \v{C}erenkov resonant mode with longitudinal waves does not suffer
any constraint in this parallel configuration, but the growth rate now
becomes significantly smaller than that of the oblique mode (when the
latter applies), by a factor $\sin^{2/3}\theta_B$.

Concerning the Whistler modes, the condition $\vert\delta\vert\,<\,1$
now leads to $Y\,\gg\,(m_e/m_p)^{-2}(\Gamma_{\rm
  sh}\sin\theta_B)^{-2}X^4$. One recovers the condition expressed in
the case of superluminal shock waves for $\sin\theta_B^2\,\sim\,1$, as
one should. At a fixed value of $\Gamma_{\rm sh}\sin\theta_B$ however,
the condition appears slightly different. Note that in the particular
case of parallel shock waves, there is another non-trivial condition
for these Whistler modes, which is related to the fact that the
eigenfrequency $\omega_{\rm Wh.}\,\simeq\,\omega_{\rm ce}\sin\theta_B$
(at maximal growth) should exceed $\omega_{\rm ci}$, in order for the
Whistler branch to apply. This translates into $X\,\ll\,\Gamma_{\rm
  sh}\sin\theta_B\,<\,1$.

Finally, the Bell non-resonant instability requires the background
protons to be magnetized, as noted in Section~\ref{sec:bell}, which
corresponds to $Y\,\gg\,\xi_{\rm cr}(m_p/m_e)^6X^6$.

\begin{figure}
\includegraphics[width=0.5\textwidth]{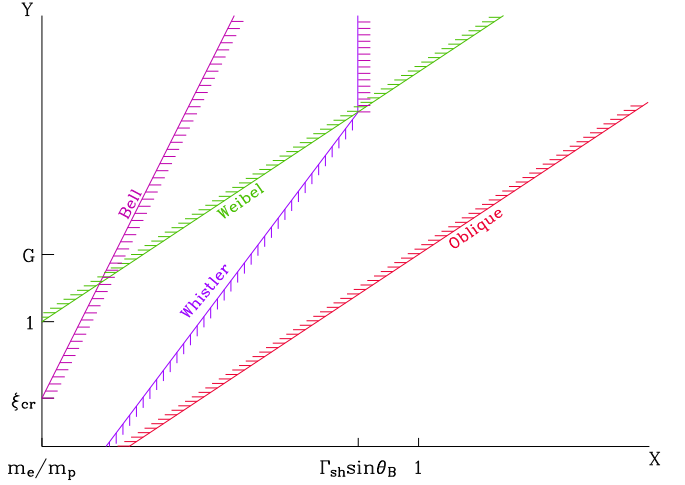}
\caption{Same as Fig.~\ref{fig:XY_perp} for the case of a parallel
  relativistic shock wave. The regions of growth are drawn at a fixed
  value of $\Gamma_{\rm sh}\sin\theta_B$ (here taken to be 0.3). The
  axes are plotted in log-log on arbitrary scale. The longitudinal
  mode can grow in all parameter space (in the idealized limit of a
  fully coherent upstream magnetic field) due to the divergence of the
  precursor length (see main text).
\label{fig:XY_par}}
\end{figure}

These various conditions are expressed in Fig.~\ref{fig:XY_par}, which
is the analog of Fig.~\ref{fig:XY_perp} for parallel shock waves. Here
as well, the limit of a pair shock can be obtained simply by taking
the limit $m_e/m_p\rightarrow 1$ and discarding the Whistler branch as
well as the Bell instability (which requires a net current to exist
upstream).

As for the oblique shock wave, it is instructive to compare the
present results to the simulations of Sironi \& Spitkovsky (2009), who
find in particular that instabilities can be triggered in parallel
shocks for a magnetisation $\sigma_{\rm SS}=0.1$. Here $\sigma_{\rm
  SS}$ corresponds to the definition of the magnetisation given in
Sironi \& Spitkovsky (2009), or to our defintion of downstream
magnetisation up to a factor $3/4$, the latter factor of $3/4$
representing the difference between enthalpy and energy density for a
relativistic gas. For a parallel shock wave, this thus corresponds to
an upstream magnetisation $\sigma_{\rm u}\,=\,0.05\Gamma_{\rm sh}^2$,
hence to $Y\,\simeq\,0.05\Gamma_{\rm sh}^4\xi_{\rm
  cr}^{-1}\,\simeq\,10^4\xi_{\rm cr}^{-1}$ for $X\,\simeq\,20$. One
may note that at this large level of magnetisation, one has
$\omega_{\rm ce}>\omega_{\rm pe}$. One can check immediately, using
the above, that neither the oblique nor the Weibel instabilities can
grow, as their respective non-magnetisation conditions are not
satisfied. The growth of instabilities observed in the simulations of
Sironi \& Spitkovsky (2009) is thus likely due to the \v{C}erenkov
resonance with the longitudinal modes that propagate along the
magnetic field, which does not suffer from these constraints.

The above also allows to understand the abrupt transition as a
function of shock obliquity: when $\sin\theta_B\rightarrow
1/\Gamma_{\rm sh}$, the growth of fluctuations is suddenly inhibited
and so is Fermi acceleration. Indeed, as $\sin\theta_B\rightarrow
1/\Gamma_{\rm sh}$, the fraction of particles that can escape to
upstream infinity vanishes, hence the precursor length now rapidly
decreases to the value given by Eq.~(\ref{eq:lfb}). This prevents the
growth of fluctuations at high magnetisation levels, as discussed
above for superluminal shock waves, and consequently this prevents
successful Fermi acceleration (Lemoine, Pelletier \& Revenu 2006).

\section{Triggering Fermi acceleration}\label{sec:Fermi}
It is important to underline that Fig.~\ref{fig:XY_perp} indicates
whether instabilities triggered by the first generation of cosmic rays
returning upstream have time to grow or not. If these instabilities
cannot be triggered by the first generation, meaning if the shock wave
characteristics are such that $(X,Y)$ lie above the thick solid line
of Fig.~\ref{fig:XY_perp}, then instabilities cannot be triggered,
either upstream or downstream (at least in the frame of our
approach). Consequently Fermi cycles will not develop, at least for
superluminal shock waves, in accordance with the arguments of Lemoine,
Pelletier \& Revenu 2006, Pelletier, Lemoine \& Marcowith 2009 and
with the simulations of Niemiec, Ostrowski \& Pohl 2006.

If, however, the initial values of $X$ and $Y$ are such that
instabilities can develop, Fig.~\ref{fig:XY_perp} suggest that these
instabilities will develop upstream and be transferred
downstream. Fermi cycles may then develop provided the appropriate
conditions discussed in Lemoine, Pelletier \& Revenu (2006) and
Pelletier, Lemoine \& Marcowith (2009) are satisfied. These conditions
have been discussed under the assumption of isotropic short scale
magnetic turbulence, and we restrict ourselves to this assumption in
the present work as well. It would certainly be interesting to
generalize this discussion to more realistic turbulence
configurations, as in Hededal et al. (2004), Dieckmann, Drury \&
Shukla (2006) for instance. However, this clearly becomes more model
dependent in terms of turbulence configuration and for this reason, we
postpone such a study to future work.

Let us discuss first the case of upstream turbulence. When particles
are scattered off short scale $\ell_{\rm c}$, but intense magnetic
fluctuations, the scattering frequency of a relativistic particle of
momentum $p$ is
\begin{equation}
\label{NUS}
\nu_{\rm s} \,\sim\, c \frac{e^2\langle \delta B^2 \rangle
}{p^2}\ell_{\rm c}\ .
\end{equation}
Since the oblique mode dominates over the Whistler and Weibel waves
over most of the parameter space, one cannot ignore the influence of
short scale electrostatic fields. These electrostatic waves lead to a
second order Fermi process in the upstream medium, with a concomittant
pitch angle scattering. Indeed, the particle scatters against random
electric fields $\pm E_\parallel$ along the shock normal ($\mathbf{x}$
direction), gaining momentum $\Delta p_\parallel\,\simeq\, \pm
eE_\parallel \Delta t$, with $\Delta t\,\simeq\,\omega_{\rm p}^{-1}$
at each interaction, and similarly in the perpendicular direction. The
initial pitch angle of the particle (with respect to the shock normal)
$\theta\,\ll\,1$ in the upstream frame, and the particle is overtaken
by the shock wave whenever $\theta\,\gtrsim\,1/\Gamma_{\rm sh}$
(Achterberg et al. 2001). This pitch angle diffuses according to:
\begin{equation} 
  \frac{\langle\Delta \theta^2\rangle}{\Delta
    t}\,\simeq\, \frac{\langle\Delta p^2\rangle}{p^2 \Delta t}
  \,\simeq\, e^2 \frac{E_\perp^2 +
    2\theta^2E_\parallel^2}{p_\parallel^2}\tau_{\rm c} \ ,
\end{equation}
for a correlation time $\tau_{\rm c}=\ell_{\rm c}/c \sim \omega_{\rm
  pe}^{-1}$.  Therefore we obtain a scattering rate similar to the
previous one (\ref{NUS}) in which the magnetic field fluctuation is
replaced by the electric field fluctuation:
\begin{equation}
\label{eq:diffrate}
\nu_{\rm s}' \sim c \frac{e^2\langle \delta E^2 \rangle}{p^2}
\ell_{\rm c}\ .
\end{equation}
This correspondence justifies that we treat the short scale electric
and magnetic fields on a similar footing and consider the total
electromagnetic energy content. A conversion of a fraction of the
energy of the beam into magnetic or electrostatic fluctuations is
expected with $\xi_{\rm em} < \xi_{\rm cr}$, with typically $\xi_{\rm
  cr} \sim 10^{-1}$ and $\xi_{\rm em} \sim 10^{-2}-10^{-1}$ (Spitkovsky
2008a). Scattering in the short scale electromagnetic turbulence will
govern the scattering process if it leads to $\langle\Delta
p^2\rangle/p^2\,\sim\,1/\Gamma_{\rm sh}^2$ on a timescale $r_{{\rm
    L}\vert B}/(\Gamma_{\rm sh}c)$, with $r_{{\rm L}\vert B}$ the
Larmor radius of first generation cosmic rays as measured upstream
relatively to the background magnetic field (see the corresponding
discussion in Pelletier, Lemoine \& Marcowith 2009). If this short
scale turbulence governs the scattering process, then Fermi
acceleration will operate. Assuming $\ell_{\rm c}=c/\omega_{\rm pe}$,
this condition amounts to:
\begin{equation}
  \xi_{\rm em} > \Gamma_{\rm sh} \left(\frac{m_p}{m_e}\right)^{1/2}
  \sigma_{\rm u}^{1/2} \ .
\label{eq:fup1}
\end{equation}
Using the fact that $\xi_{\rm em}\,<\,\xi_{\rm cr}$, this constraint
can be rewritten as a bound on $\sigma_{\rm u}$:
\begin{equation}
\sigma_{\rm u}\,\ll\, \xi_{\rm cr}^2 \frac{m_e}{m_p}\Gamma_{\rm
  sh}^{-2}\ .\label{eq:fup}
\end{equation}
This limit is very stringent indeed; in terms of our above parameters,
it can rewritten as $Y\,\ll\,X\xi_{\rm cr}/\Gamma_{\rm sh}$. We will
discuss the applicability of this inequality in concrete cases in the
following sub-section.

If this condition is not verified, the background unamplified magnetic
field remains the main agent of particle scattering upstream. In this
case, Fermi acceleration cycles can develop only if short scale
turbulence govern the scattering downstream of the shock wave. As
discussed in Pelletier, Lemoine \& Marcowith (2009), this requires:
\begin{equation}
\ell_{\rm c\vert d}\,<\,r_{\rm L\vert d}\,<\,\frac{\delta B_{\vert\rm
    d}}{B_{\vert\rm d}}\ell_{\rm c\vert d}\ ,\label{eq:fdown}
\end{equation}
where all quantities should be evaluated in the downstream rest frame,
and $r_{\rm L\vert d}$ refers to the Larmor radius of the accelerated
particles in this frame. This double inequality amounts to requiring
that $\ell_{\rm c\vert d}/c \,<\ \tau_{\rm s} \,<\ \tau_{\rm L,0}$,
i.e. that the scattering time $\tau_{\rm s}=\nu_{\rm s}^{-1}$ be
shorter than the Larmor time in the mean field $\tau_{\rm L,0}$ in
order to break the inhibition constraint of the mean field that tends
to drag the particles in the downstream flow. The scattering must also
develop in a special regime where the correlation time $\ell_{\rm
  c\vert d}/c$ is shorter than the Larmor time.  Regarding $\ell_{\rm
  c\vert d}$, two main spatial scales are to be envisaged: the
previous upstream electron skin depth, if one assumes that the typical
scale of transverse fluctuations is preserved through shock crossing,
and the downstream electron skin depth, if reorganization takes place
through shock crossing. Assuming a typical electron temperature $\sim
\Gamma_{\rm sh}m_p c^2$ behind the shock, and accounting for shock
compression of the electron density, this latter scale can actually be
written as $c\,\omega_{\rm pi}^{-1}$ ($\omega_{\rm pi}$ the {\em
  upstream} ion plasma frequency), a factor $43$ larger than the
previous one. One should also envisage the possibility that the
turbulence spectrum evolves to larger scales with time (Medvedev et
al. 2005; Lemoine \& Revenu 2006; Katz, Keshet \& Waxman 2007) but we
will not do so here. Let us consider the above two possibilities in
turn.

If $\ell_{\rm c\vert d}\,=\,c/\omega_{\rm pe}$ (upstream electron skin
depth), then the first inequality in Eq.~(\ref{eq:fdown}) can be
rewritten as $\xi_{\rm em}\,<\,m_p/m_e$ and is therefore always
satisfied. The second inequality amounts to $\sigma_{\rm d}\,<\,
(m_e/m_p)\xi_{\rm em}^2$, hence $Y\,<\,\Gamma_{\rm sh}X \xi_{\rm
  em}^2/\xi_{\rm cr}$. This latter inequality is much more
stringent. If satisfied, it means that the downstream short scale
turbulence governs the scattering process, in particular it allows the
particle to escape its orbit around the shock compressed background
magnetic field on a timescale smaller than the Larmor time in this
field. This is a necessary condition for successful Fermi cycles.

If $\ell_{\rm c\vert d}\,=\,c\,\omega_{\rm pi}^{-1}$ (equivalently,
the downstream electron skin depth), then the first inequality in
Eq.~(\ref{eq:fdown}) becomes $\xi_{\rm em}\,<\,1$, which is always
true. The second inequality reads $\sigma_{\rm d}\,<\,\xi_{\rm em}^2$
(or $Y\,<\,\Gamma_{\rm sh}^2\xi_{\rm em}^2/\xi_{\rm cr}$). We will
summarize the two above two possible cases for $\ell_{\rm c\vert d}$
and parameterize the uncertainty on $\ell_{\rm c\vert d}$ by writing
the condition as:
\begin{equation}
\sigma_{\rm d}\, \ll\, \sigma_{*} \equiv \kappa\, \xi_{\rm em}^2\ ,
\label{eq:fdown2}
\end{equation}
with $\ell_{\rm c\vert d}\,=\,\kappa c/\omega_{pi}$ and
$m_e/m_p\,\lesssim\,\kappa\,\lesssim\,1$. One should however recall
that the typical scale of electromagnetic fluctuations could evolve
with the distance to the shock front, as envisaged in Medvedev et
al. (2005), Lemoine \& Revenu (2006) and Katz, Keshet \& Waxman
(2007). This amounts to making $\kappa$ be a growing function of the
energy taking values larger than 1. The above result clearly reveals
the need for dedicated PIC simulations of shock wave at moderate
magnetisation, with realistic proton to mass ratio and geometry in
order to reduce this large uncertainty on $\kappa$ and determine the
precise conditions under which Fermi acceleration can take place.

To summarize this discussion, we obtain the following conditions for
successful Fermi acceleration. If Eq.~(\ref{eq:fup}) is satisfied [or,
  to be more accurate, Eq.~(\ref{eq:fup1})], then Fermi acceleration
will operate, because the short scale fluctuations produced upstream
are sufficiently intense to govern the scattering. In this case, it is
important to stress that Eq.~(\ref{eq:lfb}), which defines the length
of the precursor, no longer applies. It should be replaced by
Eq.~(\ref{eq:lfu}), which is larger. Physically, the precursor widens,
giving more time for the fluctuations to grow, thus reaching a higher
efficiency in terms of $\xi_{\rm em}/\xi_{\rm cr}$.  If
Eq.~(\ref{eq:fup}) is not satisfied, e.g. because the upstream
magnetization is not small enough, particles gyrate in the background
magnetic field before experiencing the short scale turbulence. Then
Fermi acceleration will operate if Eq.~(\ref{eq:fdown2}) is verified.
Consequently, a sufficient condition for the development of Fermi
cycles is $\xi_{em} > \left(\sigma_{\rm crit}/\kappa\right)^{1/2}$, or
equivalently $\sigma_* > \sigma_{\rm crit}$, where $\sigma_{\rm crit}$
is the maximum value of the upstream magnetization that allows
turbulence to grow upstream and then be transferred downstream. As
shown previously for a superluminal configuration, $\sigma_{\rm crit}
= \Gamma_{\rm sh}^{-3}\xi_{\rm cr} m_p/m_e$ for an electron-proton
plasma in the realistic case where $\Gamma_s \lesssim 150\xi_{\rm
  cr}^{1/3}$ so that the transition is governed by the excitation of
whistler waves; or $\sigma_{\rm crit}=\Gamma_{\rm
  sh}^{-2}\xi_{cr}^{2/3}(m_p/m_e)^{1/3}$ for an electron-positron
plasma, with the development of the oblique two stream instability.
The spectral index and the maximal energy remain to be determined
however. In this respect, we note that Eq.~(\ref{eq:fdown}) provides
an upper bound for this maximal energy: $\epsilon_{\rm max} \simeq
\Gamma_{\rm sh} m_pc^2 (\sigma_*/\sigma_{\rm u})^{1/2}$ in the front
frame.

The more likely development of the Fermi process is thus hybrid, in
the sense that it is of drift type upstream and of diffusive type
downstream.  As Fermi cycles develop, particles are accelerated beyond
the energy $\Gamma_{\rm sh}^2m_pc^2$ considered here for the first
generation. Although they are less numerous, they stream farther ahead
of the shock and are therefore liable to induce stronger
amplification.  One can only speculate about these issues, since the
spectral index depends strongly on the assumption made on the shape of
the turbulence spectra, upstream as well as downstream. In particular,
if the magnetic field amplified downstream through the Weibel
instability decays on scales of order of tens or hundreds of electron
inertial lengths $\delta_{\rm e}$, the particles will likely escape
towards downstream because of the lack of scattering agents, thereby
cutting off the Fermi process prematurely. Nevertheless, assuming for
the sake of discussion that Fermi cycles develop with a spectral index
$s\,\sim\,2-3$, the number density of cosmic rays streaming upstream
scales as $n_{{\rm b}\vert\rm u}(>p_*)\,\propto\, (p_*/p_0)^{1-s}$,
with $p_0\,\sim\,\Gamma_{\rm sh}^2m_pc^2$. The beam plasma frequency,
which controls the growth rates of the instabilities, $\omega_{{\rm
    p}*}(>p_*)\,\propto\,(p_*/p_0)^{-s/2}$, whereas the precursor
length $\ell_{\rm F\vert\rm u}(>p_*)\,\propto\,(p_*/p_0)$. Since the
growth rates of the resonant instabilities which develop upstream
scale as $\omega_{{\rm p}*}^{2/3}$, $s<3$ would guarantee that the
growth factor of the instabilities triggered by these high energy
particles exceeds that for the first generation.  These findings seem
in agreement with the numerical simulations of Keshet et al. (2009)
and Sironi \& Spitkovsky (2009) who observe wave growth farther from
the shock from high energy particles, as time increases.

\subsection{Applications}\label{sec:appl}
It is interesting to situate the relativistic shock waves of physical
interest in the above diagram. Here we consider three proto-typical
cases: a pulsar wind, a gamma-ray burst external shock waves expanding
in the interstellar medium, and a gamma-ray burst external shock wave
propagating along a density gradient in a Wolf-Rayet wind. We find the
following:

\begin{itemize} 

\item Pulsar winds: with $\Gamma\,\simeq\,10^6$ and $\sigma_{\rm
  u}\,\simeq\,0.01$, one finds $(X,Y)\,\sim\,(500,10^{10}\xi_{\rm
  cr}^{-1})$; the level of magnetization is thus so high that no wave
  can grow, either upstream or downstream. Fermi acceleration should
  consequently be inhibited.

\item Gamma-ray burst external shock waves expanding in the
  interstellar medium: for $\Gamma\,\simeq\,300$ and $\sigma_{\rm
    u}\,\sim\,10^{-9}$ (i.e. $B\,\sim\,3\,\mu$G), one finds
  $(X,Y)\,\sim\,(0.1,10^{-5}\xi_{\rm cr}^{-1})$. Wave growth should be
  efficient both usptream and downstream. Concerning Fermi
  acceleration, Eq.~(\ref{eq:fup}) amounts to $Y\,<\,\xi_{\rm
    cr}m_e/m_p$. It can thus be only marginally satisfied. However,
  Eq.~(\ref{eq:fdown2}) is most likely satisfied, so that Fermi
  acceleration should develop, even in the early afterglow phase when
  $\Gamma_{\rm sh}\,\sim\,300$.

\item Gamma-ray burst external shock waves propagating along a density
  gradient in a Wolf-Rayet wind: taking a surface magnetic field of
  $1000\,G$ for a $10R_\odot$ Wolf-Rayet progenitor, the magnetization
  at distances of $10^{17}\,$cm is $\sigma_{\rm u}\,\sim\,10^{-4}$
  (Crowther 2007). This gives $(X,Y)\,\sim\,(0.1, \xi_{\rm
    cr}^{-1})$. Growth may or may not occur in this case, depending on
  the precise values of $\Gamma_{\rm sh}$, $\sigma_{\rm u}$ and
  $\xi_{\rm cr}$. In detail, the condition for Weibel growth
  $Y\,\lesssim\,1$ is likely not verified for the above fiducial
  values, but could be verified in less magnetized winds and at later
  stages of evolution, with a smaller value of $\Gamma_{\rm sh}$. The
  condition for growth of Whistler waves, $Y\,\lesssim\,1/X$, may be
  satisfied if $\xi_{\rm cr}\,\gtrsim\,0.1$ and it is likely to be
  more easily verified at smaller values of $\Gamma_{\rm sh}$ and
  $\sigma_{\rm u}$. Finally, the (most stringent) condition for growth
  of the oblique mode, Eq.~(\ref{eq:obllimmag}), is likely not
  verified in the initial stages with $\Gamma_{\rm sh}\simeq300$ and
  the above fiducial value of $\sigma_{\rm u}$, but would be verified
  if $\sigma_{\rm u}$ were smaller.

However, Eq.~(\ref{eq:fup}) cannot be satisfied in this case, meaning
that the orbit of the particle upstream is governed by the wind
magnetic field, not by the amplified short scale component. Regarding
the bound Eq.~(\ref{eq:fdown2}), it can be satisfied, depending on the
values of the wind magnetisation and most particularly on the value of
$\kappa$. The possibility of Fermi acceleration thus remains open in
this case. More work is necessary to understand the properties of
downstream turbulence in order to determine whether particle can
eventually be accelerated.

\end{itemize}


\subsection{Further considerations}

It is important to emphasize that we do not understand yet the
structure of a relativistic shock front in detail. In the previous
section we have assumed that the shock front is structured like a
non-relativistic front and just extended the non-relativistic
results. Since MHD compressive instability and extraordinary ionic
modes can be excited, we cannot exclude that the foot be full of
relativistically hot protons and electrons of similar temperature
$\bar \gamma m_pc^2$ with $1\, \ll \, \bar \gamma\,\leq \,\Gamma_s$. 
In that case the plasma
response would be different, because the intermediate whistler range
(and also extraordinary range) would disappear so that the plasma
would behave like a relativistic pair plasma. Then, the relevant
instabilities are the Weibel and oblique modes (in the unmagnetized
approximation). The length of the precursor and the Weibel growth rate
remain unchanged, hence the domain of growth of the Weibel instability
also remains unchanged. The growth rate of the oblique mode is however
reduced because the background plasma frequency is smaller by a ratio
$(\gamma m_p/m_e)^{1/2}$. Therefore the condition of growth on the
advection timescale now reads $Y\,\ll\,\xi_{\rm cr}^{-1/3} \Gamma_{\rm
  sh}^{1/3} X^{-1/3} (\gamma m_p/m_e)^{-1/3}$. The ratio of the growth
rates of the oblique mode to the Weibel mode can be written as
$(\gamma \xi_{\rm cr})^{-1/6}$, hence the Weibel instability becomes
the dominant mode if $\gamma\,\gg\,\xi_{\rm cr}^{-1}$.

In the downstream plasma, the magnetic fluctuations generated by the
Weibel instability are expected to disappear rapidly because they do
not correspond to plasma modes. Whistler and other resonant eigenmodes
(when they are excited) are however transmitted and although they are
not excited downstream, their damping is weak.  When Fermi cycles
develop, they create ``inverted'' distribution downstream, that should
produce a maser effect. 

Tangled magnetic field carried by the upstream flow are very
compressed downstream and thus opposite polarization field lines come
close together. This should produce magnetic reconnections in an
unusual regime where protons and electrons have a similar relativistic
mass of order $\Gamma_{\rm sh} m_pc^2$. Such a regime of reconnection
deserves a specific investigation with appropriate numerical
simulations. Despite magnetic dissipation, reconnections would
probably create a chaotic flow that favors diffusion of particles from
downstream to upstream. \\

\section{Conclusions}

In this work, we have carried out a detailed study of the
micro-instabilities at play in the precursor of a ultra-relativistic
shock wave. The main limitation for the growth of these waves is
related to the length of precursor, which is itself related to the
level of magnetisation in the upstream plasma (where magnetisation
refers to the background field, not the shock generated short scale
fields).  Nevertheless, we have found electronic and ionic
instabilities that grow sufficiently fast in the precursor of a
relativistic shock. The fastest growing instabilities are due to the
\v{C}erenkov resonance between the beam of accelerated (and shock
reflected protons) and the upstream plasma Whistler waves and
electrostatic modes. The Weibel instability, which is non-resonant by
essence, is also excited, but its growth is generally superseded by
that of the previous modes. The strongest amplification occurs on very
short spatial scales $\sim\delta_{\rm e}$, the electron skin depth in
the upstream plasma. Our results are summarized in
Fig.~\ref{fig:XY_perp} for the generic case of relativistic
superluminal shock waves, which delimits the domains in which
electromagnetic modes are excited in terms of shock Lorentz factor and
upstream magnetisation and defines the critical value $\sigma_{crit}$
of the magnetization below which Fermi process can operate.  Figure
\ref{fig:XY_par} presents the corresponding limitations for the case
of parallel shock waves; in this case, the growth of instabilities is
made much easier by the divergence of the precursor length for a
fraction of the particles returning upstream. Our results explain some
features of recent PIC simulations of relativistic pair shocks of
various geometries and magnetisation levels.

We have discussed the conditions under which Fermi acceleration can
proceed superluminal shock waves once a significant fraction of the
cosmic ray energy has been dumped into these short scale
electromagnetic fluctuations.  Fermi acceleration can operate if the
upstream magnetisation ($\sigma_{\rm u}$) or downstream magnetisation
($\sigma_{\rm d}$) are low enough for the shock generated turbulence
to govern the scattering of particles. This is the second condition
that states the required level of electromagnetic energy density
versus the magnetization. This requires either $\sigma_{\rm u}\,\ll\,
\xi_{\rm em}^2 (m_e / m_p)\Gamma_{\rm sh}^{-2}$ (for upstream
scattering), which is however difficult to fulfill or $\sigma_{\rm
  d}\,\ll\, \kappa\, \xi_{\rm em}^2$ (for downstream scattering, which
is easily fulfilled with a low level of turbulence); $\xi_{\rm em}$
indicates the fraction of incoming energy transferred into
electromagnetic fluctuations, with $\xi_{\rm
  em}\,\sim\,10^{-2}-10^{-1}$ generally indicated by PIC simulations,
and $\kappa$ is a fudge factor that encaptures our ignorance of the
transfer of electromagnetic modes excited upstream through the shock,
$m_e/m_p\,\lesssim\,\kappa\,\lesssim\,1$ (and it may be even larger
depending on the particle energy if the scale of the electromagnetic
flucutations evolves with the distance to the shock). We emphasize the
need for PIC simulations with realistic geometry, realistic proton to
electron mass ratios and moderate magnetisation (of order of the
above) in order to lift this uncertainty on $\kappa$ and to determine
the precise conditions under which Fermi acceleration can take place.
This limitation also places a strict upper bound on the maximum energy
that is achievable through the Fermi process, namely $\epsilon_{\rm
  max} \simeq \Gamma_{\rm sh}m_p c^2 (\sigma_{\rm crit}/\sigma_{\rm
  u})^{1/2}$, with $\sigma_{\rm crit}$ the maximum magnetisation that
allows waves to grow, and $\sigma_{\rm u}$ the upstream background
magnetisation (see discussion after Eq.~(\ref{eq:fdown2}). Beyond this
intrinsic limit, the scattering time indeed becomes longer than the
Larmor time in the mean field downstream, so that the particle is
advected downstream by the mean field and Fermi cycles end.

We have also applied our calculations to several cases of
astrophysical interest.  In practice, we find that terminal shocks of
pulsar winds have a magnetisation level that is too high to allow for
the amplification of short scale electromagnetic fields, so that
particle acceleration must be inhibited. We have found that gamma-ray
burst external shock waves propagating into a typical interstellar
medium should lead to strong amplification of the magnetic field and
to Fermi cycles, even at high Lorentz factor. The energies reached by
the suprathermal electrons can easily explain the afterglow emission
through jitter radiation (Medvedev 2000). However, if the shock wave
propagates in a stellar wind, the upstream magnetisation may be too
large to allow for particle acceleration, eventhough magnetic field
amplification should take place.

\section*{Acknowledgments}
We warmly acknowledge A. Marcowith for collaboration at an earlier
stage of this work and for a careful reading of the manuscript. We
also thank the referee for a detailed reading of the manuscript and
for useful suggestions. One of us (G.P.) acknowledges fruitful
discussions with J. Arons, A. Bell, L. Drury, J. Kirk, Y. Lyubarsky,
J. Niemiec, M. Ostrowski, B. Reville, H. V\"olk, A. Spitkovsky and
L. Sironi. We also thanks an anonymous referee for his accurate and
detailed analysis of our work.

\end{document}